\documentclass[%
 reprint,
amsmath,amssymb,
aps,
prx,
reprint,
floatfix,
showkeys,
superscriptaddress]{revtex4-2}

\usepackage[english]{babel}

\usepackage{graphicx}
\usepackage{dcolumn}
\usepackage{bm}
\usepackage{amssymb}   
\usepackage{subcaption}
\usepackage[export]{adjustbox}
\usepackage{amsmath}
\usepackage{ragged2e}
\usepackage{xcolor}

\begin{document}

\preprint{APS/123-QED}

\title{How Geometry Tames Disorder in Lattice Fracture}

\author{Matthaios Chouzouris}
\email{m.chouzouris@sms.ed.ac.uk}
\affiliation{Institute for Infrastructure and Environment, School of Engineering, University of Edinburgh, Edinburgh EH9 3FG, UK}

\author{Leo de Waal}
\affiliation{Institute for Infrastructure and Environment, School of Engineering, University of Edinburgh, Edinburgh EH9 3FG, UK}
\affiliation{Faculty of Engineering Sciences, University College London, London WC1E 7JE, UK}

\author{Antoine Sanner}
\author{Alessandra Lingua}
\author{David S. Kammer}
\affiliation{Institute for Building Materials, ETH Zurich, 8093 Zurich, Switzerland}

\author{Marcelo A. Dias}%
\email{mdias@ed.ac.uk}
\affiliation{Institute for Infrastructure and Environment, School of Engineering, University of Edinburgh, Edinburgh EH9 3FG, UK}

\makeatletter
\renewcommand\frontmatter@affiliationfont{\small}  
\makeatother


\date{\today}

\begin{abstract}
We investigate the fracture behavior of pre-cracked triangular beam-lattices whose elements have failure stresses drawn from a Weibull distribution. Through a statistical analysis and numerical simulations, we identify and verify the existence of three distinct failure regimes: (i) disorder is effectively suppressed, (ii) disorder manifests locally near the crack tip, modifying the crack morphology, and (iii) disorder manifests globally, leading to initially diffuse failure. Our model naturally reveals the key parameters governing this behavior: the Weibull modulus, quantifying the spread in failure thresholds, and a geometric quantity termed the \textit{Slenderness Ratio}. We also reproduce the disorder-induced toughening reported in previous experimental and numerical studies, further demonstrating that its manifestation depends non-monotonically on disorder. Crucially, our results indicate that this toughening cannot be simply connected to the amount of damage in the lattice, challenging interpretations that attribute increased fracture energy solely to enhanced crack tortuosity or diffuse failure. Overall, our results establish geometry as a powerful control parameter for regulating how disorder is expressed during fracture in beam-lattices, with broader implications for the disorder-induced toughening in engineered materials.

\end{abstract}

\maketitle

\section{\label{sec:level1} Introduction}

Mechanical metamaterials are engineered structures whose unusual mechanical properties derive mainly from their geometry and topology, rather than their chemical composition. Often realized as lattice-based structures, these materials can exhibit extraordinary behaviors, such as bespoke stiffnesses and negative Poisson's ratios \cite{ZhengXiaoyu2014UUMM, LyuYongtao2024Anmm, SchwaigerR.2019Temo}, bistability \cite{RafsanjaniAhmad2016Bamm}, tunable stress-strain curves \cite{ZhangHang2018Smmw} and programmable unconventional damage responses \cite{de2025cracking}. Additive and subtractive manufacturing techniques have enabled the realization of these and many more intricate architectures, making previously theoretical designs accessible in practice \cite{KhosravaniMohammadReza2025Fomm}. 

Such fabrication methods inevitably introduce imperfections, thus presenting an abundance of defects and sources of disorder that are unavoidable. From unwanted porosity and surface roughness \cite{masuo2018influence}, residual stresses and geometric distortions \cite{silva2025investigation}, to microstructural defects such as notch-defects \cite{lertthanasarn2020mechanical}, these irregularities threaten to alter the theoretically engineered responses, potentially compromising the functionality of architected lattices.

Fracturing in the presence of disorder in lattice metamaterials is particularly intriguing, as inhomogeneities can fundamentally alter fracture behavior, not only affecting overall toughness but also the way cracks propagate through the structure \cite{shekhawat2013damage}. Certain forms of disorder, such as slight variations in strut stiffness or node positions, can deflect cracks, promote more evenly distributed failure, and lead to improved fracture toughness \cite{ziemke2024defect, fulco2025fracture, urabe2010fracture,fulco2025disorder,hartquist2025fracture}. Larger disorder has also been associated with reduced toughness \cite{fulco2025disorder, hartquist2025fracture}, hinting to a potential non-monotonic relation between the two. Despite these insights, the literature remains limited in strategies to systematically control or tune the mechanical expression of disorder---for example, by influencing crack-path statistics such as the total number of broken bonds and the expected fracture toughness---and most studies remain observational rather than prescriptive.

In this work, we investigate the failure characteristics of a beam-jointed triangular lattice subjected to structural inhomogeneities---a system commonly referred to as the \textit{Random Beam Model} \cite{alava2008role}. To enable more physically meaningful insights into damage evolution in architected lattices, we extend this classical model by further discretizing each beam into three elements. This refinement, consistent with approaches found in recent studies \cite{de2025architecting}, enhances the model’s ability to capture the dominant bending deformation modes of individual lattice members. As a result, it provides a more faithful representation of the system’s local distortions and overall mechanical response under loading.

Focusing on Weibull-distributed disorder on the failure stresses of its beams \cite{weibull_statistical_1939, bertalan2014fracture}, we draw conclusions about how the expression of quenched disorder depends on geometry, and specifically on the \textit{Slenderness Ratio} (SR), which is the ratio of unit cell size ($a$), to in-plane beam thickness ($t$): $\lambda \equiv a/t$. By deriving a mechanically informed statistical framework to describe the sequential failure of lattice elements, we establish quantitative relationships linking the disorder strength to key damage metrics, including the amount of damage on the main crack and the probability of diffuse/remote failure during the early stages of fracture. This analysis culminates in a type of phase diagram that delineates distinct failure regimes within the space of disorder (Weibull modulus $n$) and geometric slenderness.

We validate this theoretical framework through detailed numerical simulations, demonstrating strong agreement with observed failure behaviors. Our results further reproduce the increase in apparent fracture toughness observed in studies with different types of disorder \cite{urabe2010fracture,ziemke2024defect,fulco2025disorder, fulco2025fracture,hartquist2025fracture}, while also revealing that this enhancement depends sensitively---and non-monotonically---on disorder strength and lattice geometry. Importantly, we find that disorder-induced toughening does not admit a simple mechanistic interpretation based solely on crack-path elongation or more distributed failure, challenging certain commonly invoked explanations in the literature (e.g., \cite{urabe2010fracture, fulco2025disorder}). At the same time our findings are consistent with interpretations that attribute apparent toughening to local crack arrests arising from spatial variations in fracture resistance \cite{sanner2025less, charles2002crack, curtin1990microcrack}. Most importantly, our results show that by tuning the SR---effectively manipulating the micromechanical characteristics of the lattice---it becomes possible to control how quenched disorder manifests at the structural scale. This provides a route toward the deliberate design of lattice architectures where fracture behavior can be tailored, shifting the focus from merely predicting failure under disorder to actively engineering its expression.


\section{Problem Statement}\label{sec:Problem formulation}

\begin{figure}[htbp]
    \centering
    \includegraphics[width=\columnwidth]{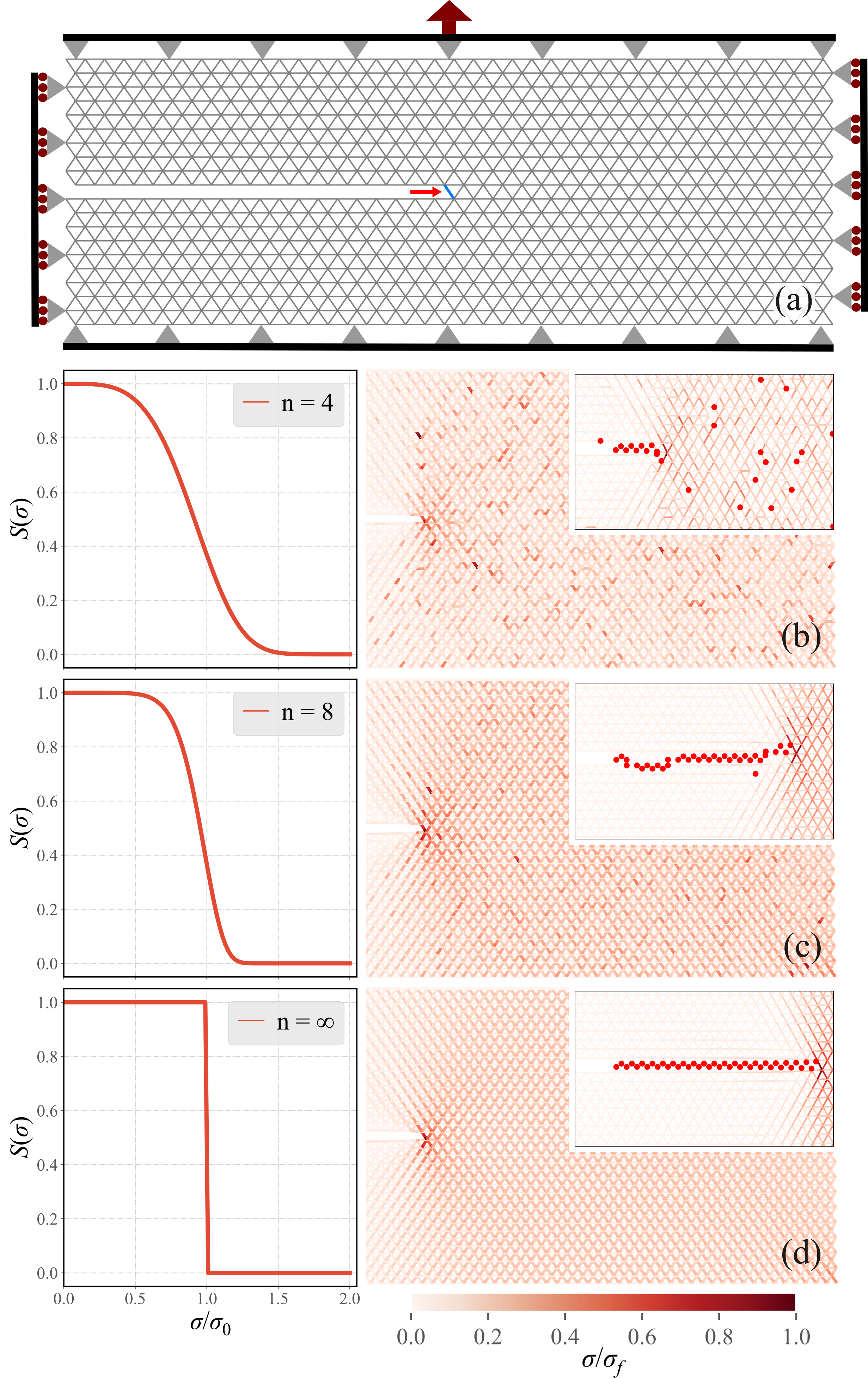} 
    \caption{\justifying Problem setup. (a): We consider rectangular domains of triangular lattices, with a half domain horizontal crack along the vertical midpoint, pinned to the horizontal, and with roller constraints along the vertical boundaries. The horizontal boundaries are kept parallel and moved apart to stretch the lattice and induce damage. The most stressed bond under these loading conditions is shown by the red arrow and colored blue. (b-d): Visualization of the lattice stresses just before and long after (insets on the top right of each) the onset of failure, in quasi-static loading. Elements in the lattice are colored according to their proximity to failure, which is the ratio of their combined axial-bending stress ($\sigma$) to their individual failure stresses ($\sigma_f$); insets on the left show the respective Weibull survival distributions. Increasing disorder (smaller $n$) progressively hides information about the elements' stresses.}
    \label{fig:Figure1}
\end{figure}

When modeling inhomogeneities in materials, several statistical approaches are possible; the Weibull distribution has emerged as one of the most widely used and effective choices, particularly for brittle failure \cite{weihull1951statistical}. The Weibull distribution can be easily understood when looking at the survival probability of a material volume $V$ with an internal stress distribution $\sigma(\boldsymbol{r})$, as a function of position $\boldsymbol{r}$ in the elastic body. Introducing two material parameters---$\sigma_0$ a characteristic failure stress, and $n$ the Weibull modulus---this survival probability takes the form:
\begin{equation} \label{eq:weibull_surv}
    S_V(\sigma(\boldsymbol{r});\ \sigma_0, n) = e^{-\frac{1}{V_0}\int_V \mathrm{d}^3\boldsymbol{r}\left(\frac{\sigma(\boldsymbol{r})}{\sigma_0}\right)^n}.
\end{equation}
Regarding the failure of a small volume (say $V_0$) with a uniform stress $\sigma$ across it, one may also write a rate of failure:
\begin{align}
    W_{V_0}(\sigma; \sigma_0, n) &= \frac{\partial}{\partial \sigma} (1 - S_{V_0}(\sigma; \sigma_0, n))\nonumber \\ 
    &= \frac{n}{\sigma_0}\left(\frac{\sigma}{\sigma_0}\right)^{n-1} e^{-\left(\frac{\sigma}{\sigma_0}\right)^n}.
\end{align}
The modulus $n$ controls the spread of failure stresses: larger values of $n$ correspond to narrower distributions and sharper cutoffs for the survival probability. In the limiting case $n \rightarrow \infty$, the distribution approaches a step function, where failure occurs deterministically once $\sigma$ exceeds $\sigma_0$ (see Fig.~\ref{fig:Figure1}(b--d) for an instructive visualization). In the uniform case the most stressed element always fails first. Inhomogeneities in the failure stresses effectively obscure information about the beam stresses, making them increasingly irrelevant. This leads to failure of elements other than the most stressed one, influencing the resulting failure pattern (see the insets of Fig.~\ref{fig:Figure1}(b--d) for examples).

Throughout this study we consider rectangular domains composed of triangular lattices containing a pre-existing horizontal crack of half-domain length, positioned along the vertical midline (see Fig.~\ref{fig:Figure1}(a)). The lattice is constrained such that the horizontal boundaries are pinned to remain parallel and are displaced quasi-statically to impose tensile loading, with roller boundary conditions along the vertical boundaries. This loading protocol induces a predominantly mode-I fracture response and allows damage to nucleate and evolve from the crack tip, with the rollers preventing sharp stress concentrations in the domain corners. Material parameters in the main text are chosen to emulate the behavior of \textit{Polymethyl Methacrylate} (PMMA)---a widespread and inherently stiff and brittle plastic---with elastic modulus, mode axial and bending failure stresses taken from~\cite{de2025cracking}, set to 3120 MPa, 45 MPa and 85 MPa, respectively.

To systematically explore the interplay between disorder and lattice geometry, we vary a single geometric control parameter, the \textit{Slenderness Ratio} (SR), while keeping the overall lattice density (element cross-sectional area) fixed. This is achieved by varying the out of plane beam thickness ($b$) accordingly \cite{de2025architecting}. We also maintain constant material parameters throughout. This choice ensures that changes in fracture behavior can be attributed to geometry rather than trivial variations in material content or stiffness. Varying SR modifies the relative importance of bending and axial deformation in the lattice beams, thereby influencing the extent to which stresses can be redistributed in the vicinity of the crack tip. As a result, SR provides a controlled way to tune the lattice’s micromechanical response and, in turn, its sensitivity to disorder, offering a complementary axis to the statistical variability introduced through the failure thresholds.

In what follows, we begin by discussing how geometry-dependent micromechanics influence the likelihood of damage in elements other than the most stressed one---hereafter referred to as \textit{anomalous} damage events---and how these events affect crack propagation and morphology (Sec. \ref{sec:micromech_anom}). We then proceed to quantify their likelihood using a statistical description of failure (Sec. \ref{sec:statistical_model}). Next, we combine the insights from these two sections to construct a mechanically informed statistical model for the damage evolution (Sec. \ref{sec:mechanicall_statistical}). We conclude our investigation by testing these predictions using a lattice fracture simulator (Sec. \ref{sec:results}).

\section{Micromechanics and Anomalous Failure}\label{sec:micromech_anom}

\begin{figure*}[!ht]
    \centering
    \includegraphics[width=0.95\textwidth, left]{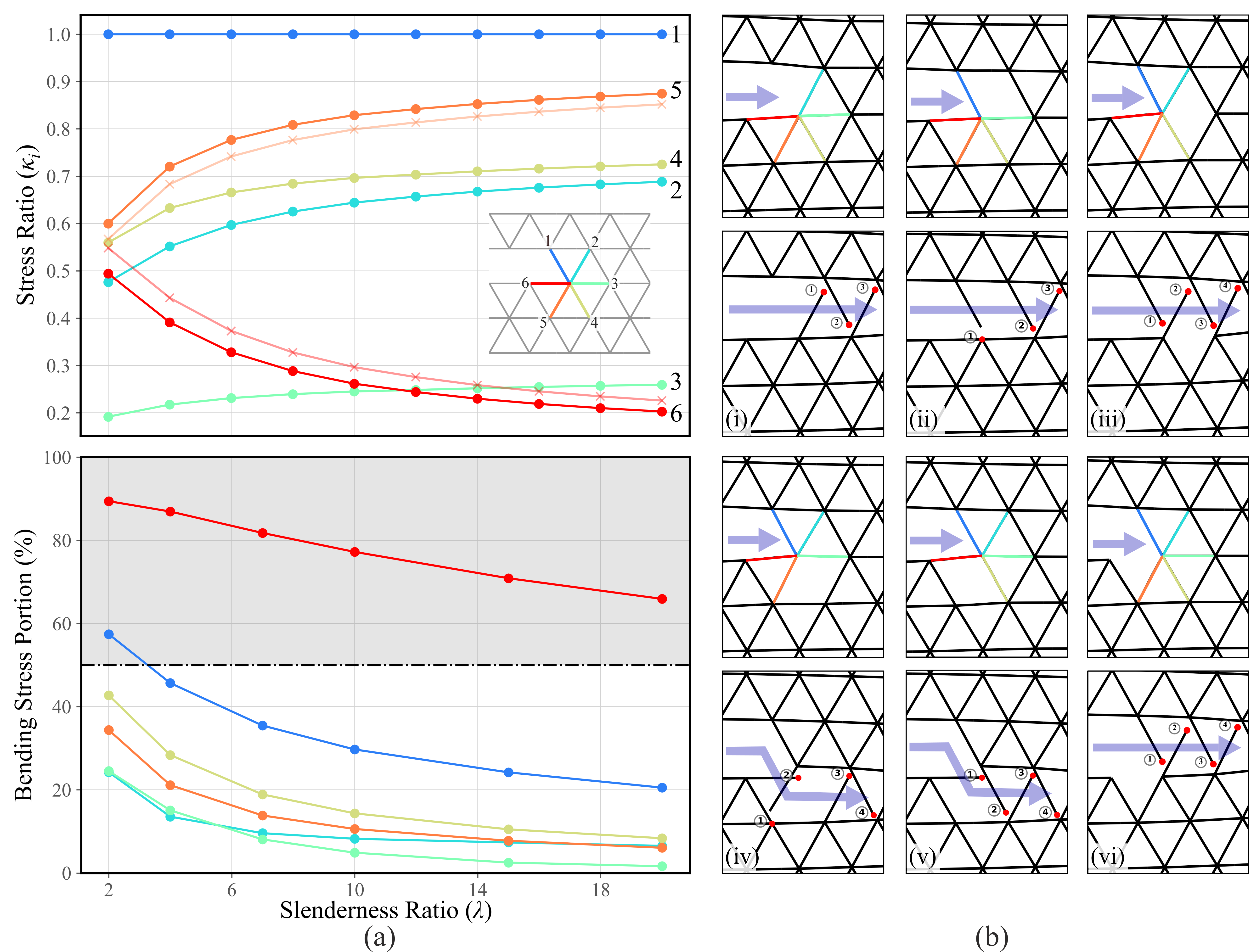} 
    \caption{\justifying Micromechanics and damage evolution after an anomalous failure event. (a): Crack-tip stress hierarchy and its geometric control. Top: Ratios of the maximum stresses in each of the six crack-tip beams to that of the most highly stressed beam (element 1, shown in blue), plotted as a function of the \textit{Slenderness Ratio} ($\lambda$). Solid lines correspond to the baseline failure criterion, with ultimate axial and bending stresses taken as $45$ MPa and $85$ MPa, respectively. Faint lines for elements 5 and 6 show the modified stress ratios obtained when the ultimate bending and axial stresses are taken to be equal (at $45$ MPa), illustrating how changes in bending failure strength alter the relative criticality of crack-tip elements. Bottom: Relative contributions of axial and bending stresses to the total stress in each crack-tip beam as a function of SR. (b): Resulting crack-path after removing each one of the \textit{crack-tip} elements, at $\lambda = 10$, and no disorder ($n \rightarrow \infty$). The red dots depict split nodes, while the circled numbers show the order of subsequent failure events. Observe that subplots iii-vi, contain one more broken element within their lower frame compared to i-ii, owing to the additional bond the crack had to remove in order to transverse the same horizontal distance. Beam thickness is not drawn to scale.}
    \label{fig:Figure2}
\end{figure*}

As already discussed, in the limit $n \rightarrow\infty$ the damage evolves by always removing the most highly stressed element in the lattice once that reaches its critical stress. This element is most clearly identified by the red arrow in Fig.~\ref{fig:Figure1}(a). Sequential removals of this element as the crack advances therefore produces a straight, horizontal crack, as illustrated in the inset of Fig.~\ref{fig:Figure1}(d).

The persistence of this failure sequence depends on the relative stresses borne by the remaining crack-tip elements. The lower panel of Fig.~\ref{fig:Figure2}(a) shows that the stresses in these beams arise from different combinations of axial and bending contributions, whose relative weights are controlled by the SR. Increasing SR suppresses bending stresses in all beams, but this suppression is non-uniform: the element with a larger bending contribution than the most stressed element experiences a faster reduction in total stress, whereas elements dominated by axial loading are less affected. As a consequence, the upper panel of Fig.~\ref{fig:Figure2}(a) shows that the ratios of crack-tip stresses relative to the most stressed element diverge or converge to it with increasing SR, depending on their bending content. In this way, SR systematically reshapes the hierarchy of crack-tip stresses even in the absence of disorder. 

The upper panel of Fig.~\ref{fig:Figure2}(a) also illustrates how modifying the relative axial and bending failure stresses for the elements, may reshape the crack-tip stress ratios. In addition to the baseline case, we show the stress ratios for elements $5$ and $6$, obtained when the ultimate bending and axial stresses are taken to be equal (set both at $45$MPa). Under this modified failure criterion, elements whose stresses are less dominated by bending become less critical relative to the most stressed element, while strongly bending-dominated elements---most notably element 6---become comparatively more critical. This highlights that, beyond controlling stress partitioning through SR, lattice beam geometry can also influence fracture behavior by tuning the relative importance of bending and axial failure modes. Since bending failure stresses can be adjusted through beam geometry \cite{timoshenko2012theory}, this provides an additional, independent avenue for regulating how disorder manifests through crack-tip micromechanics. While we do not explore this degree of freedom further here, it underscores the broader role of geometry in shaping failure pathways in disordered lattices.

The SR-dependent reordering of crack-tip stresses has direct implications once disorder is introduced. As variability in failure thresholds increases, elements whose stresses approach that of the dominant crack-tip beam become increasingly likely to fail, enabling damage events that deviate from the deterministic stress-driven sequence. In the following, we investigate how such \textit{anomalous} failure events alter crack propagation and morphology.

As shown in Fig.~\ref{fig:Figure2}(b), the removal of a crack-tip element in a uniform lattice can give rise to two qualitatively distinct outcomes. Most generally, the failure of any element other than $1$ or $2$, triggers additional damaged sites appended to the main crack, increasing the number of broken bonds required for the crack to advance a given horizontal distance (e.g., Fig.~\ref{fig:Figure2}(b,iii--vi). We refer to this process as \textit{scattering}. Scattering is almost always accompanied by a change in the lattice row along which the crack propagates (e.g., Fig.~\ref{fig:Figure2}(b,iv--v), because the elements that cause row changes when broken are also the most stressed ones (i.e., elements $4$ and $5$). Nonetheless, we consider the fundamental event increasing crack tortuosity to be the addition of bonds to the main crack; the row change is treated as a frequent, geometric consequence of scattering.

It is not hard to see that the occurrence of scattering events depends solely on which crack-tip element is removed and is independent of SR, since removing elements that would normally be involved in the horizontal propagation sequence (i.e., $1$ or $2$) does not add failed elements to the main crack. By contrast, the specific row along which the crack continues, following scattering, depends on both the removed element i, and SR. We note that this information can be encoded using a binary-valued map for each SR and each crack-tip beam, capturing which row the crack follows post failure (see Appendix \ref{sec:appendix_scattering_map} for the full map).

Now, if at any point in its propagation there is a probability $P_s$ that the crack scatters, then---given our assumptions and the geometry of a triangular lattice---the expected number of failed bonds reads 
\begin{equation}\label{eq:crack_bonds}
    N_f = 2 N_h \ (1 + P_s),
\end{equation} 
where $N_h$ is the number of unit cells the crack has traversed in the horizontal direction. The factor of two reflects the fact that two bonds are removed for each unit-cell advance of a straight horizontal crack. One can also write an associated effective (dimensionless) crack-path length
\begin{align}\label{eq:crackpath_length} 
    \tilde L\equiv \frac{L}{a N_h} = (1+P_s), 
\end{align}
with $a$ the unit cell size.

To conclude, this section has highlighted a connection between the SR and the quantity of broken bonds (or the effective crack path length), with the whole process mediated by the disorder strength $n$. This means that the SR is a degree of freedom in the geometric design space that can potentially be used to express or suppress quenched disorder during crack propagation. In the subsequent section we are going to quantify the probability of anomalous failure events from first principles, in order to eventually inform a quantitative model for the amount of excess broken bonds in a lattice as a function of SR and Weibull modulus.



\section{Statistical Modeling} \label{sec:statistical_model}
\subsection{Likelihood of Anomalous Failure Events} \label{sec:anomalousfailureprob}

Recall that an anomalous failure event is any damage event that does not break the most stressed element in the lattice. To quantify the likelihood of such events, we begin by considering two distinct elements, labeled $i$ and $j$. For concreteness, the reader may think of element $i$ as the most stressed crack-tip element (1) and of element $j$ as any other element in the crack-tip set; however, the analysis that follows is fully general and does not rely on this specific choice. Our goal is to infer the probability that element $i$ fails before element $j$ given their relative stresses. As established in the previous section, these stress ratios are known and depend explicitly on the geometric SR.

Given the chosen distribution (Weibull), the rate of element $i$ failing at stress $\sigma_i$ when at the same time element $j$ is stressed to $\sigma_j$ and still survives, is:
\begin{equation}
    g(\sigma_i, \sigma_j) = W(\sigma_i; \sigma_0, n) \ S(\sigma_j; \sigma_0, n).
\end{equation}
This expression implicitly assumes independent and identically distributed failure thresholds for $i$ and $j$, ignoring the potential for spatial correlations. Furthermore, under the assumption of linearity, we can set $\sigma_i = \kappa_{ij} \ \sigma_j$, where $\kappa_{ij}$ is a constant that only depends on the geometry (see Appendix \ref{sec:Appendix_nonlinear} for non-linear corrections). Now, given this assumption and the form of Eq. (\ref{eq:weibull_surv}), we can absorb $\kappa_{ij}$ in a re-scaling of the reference failure stress for the $i^{\textit{th}}$ beam:
\begin{equation}
    S(\kappa_{ij} \sigma_j; \sigma_0, n) = S(\sigma_j; \sigma_0/\kappa_{ij}, n).
\end{equation}
This is an important point because it leads to a useful redefinition of the rate $g(\sigma_i, \sigma_j)$, as a function of only one variable (say, $\sigma_j \equiv \sigma$ here), which implies
\begin{align}
    g(\sigma_i, \sigma_j) &\equiv g(\sigma) \nonumber \\ 
    &= \frac{n \kappa_{ij}}{\sigma_0}\left(\frac{\kappa_{ij}\sigma}{\sigma_0}\right)^{n-1} e^{-(1+\kappa_{ij}^n)\left(\frac{\sigma}{\sigma_0}\right)^n}.
\end{align}
The total probability of element $i$ breaking before $j$ will then be an integral of the rate $g(\sigma)$ over all possible stresses. Defining the random variables $s_i = \sigma_i/\sigma_{i,f}$, where $\sigma_{i,f}$ is the failure stress of element $i$, this probability is equivalently written as
\begin{equation}
    P(s_i > s_j)= \int_0^{\infty} g(\sigma) \mathrm{d}\sigma= \frac{\kappa_{ij}^n}{1+\kappa_{ij}^n} \label{eq:p_anomalous}.
\end{equation}
Evidently, the final expression is independent of $\sigma_0$, and only depends on the relative stress ratio $\kappa_{ij} = \sigma_i/\sigma_j$, and the Weibull modulus. Note that this expression is normalized, with $P(s_i < s_j) = 1 - P(s_i>s_j)$, and it evaluates to $0$ in the limit $n \rightarrow \infty$, $\forall \  \kappa_{ij}<1$. 
 
\subsection{Renormalization and the \textit{Weibull Stress}}

\begin{figure}[htbp]
    \centering
    \includegraphics[width=0.7\columnwidth]{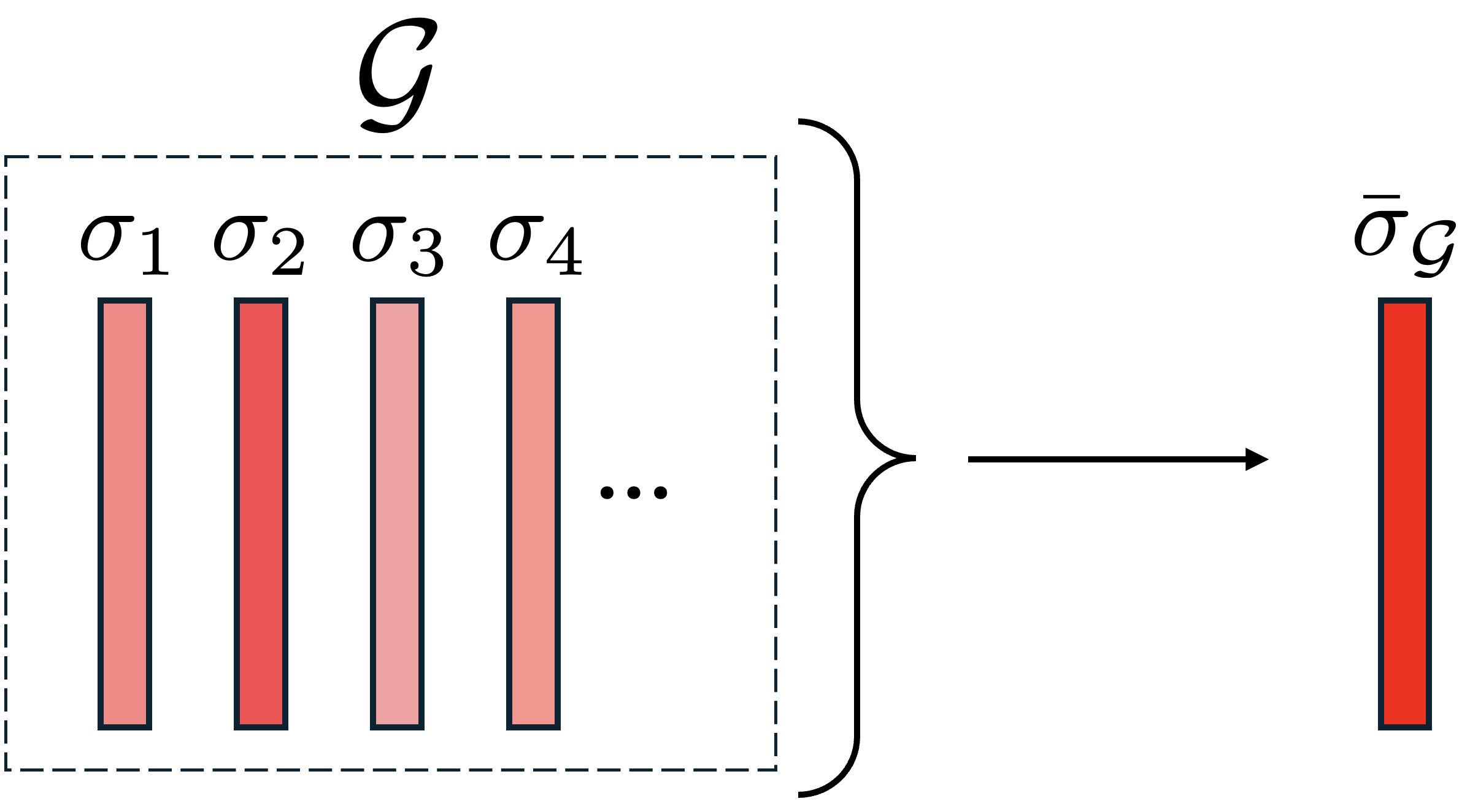} 
    \caption{\justifying Renormalization of a set $\mathcal{G}$ of $m$ elements. The resulting element can be regarded to be at a Weibull stress $\bar \sigma_{\mathrm{ \mathcal{G}}} \geq max(\{\sigma_1, \sigma_2, ... \sigma_m \})$, i.e., larger than any of the stresses of its constituents. The resulting element survives \textit{iff} all of its constituents also survive.}
    \label{fig:Figure3}
\end{figure}

We are further interested in extending this pairwise comparison to the likelihood that failure occurs within one group of elements before another. This question is, for example, directly relevant for quantifying the probability of diffuse failure, which is defined here as failure occurring at any location other than the six crack-tip elements. To that end, a useful property of the Weibull distribution is that the survival probability of a set $\mathcal{G}$ of $m$ elements, with stresses $\{\sigma_1, \sigma_2, ... \sigma_m \}$, retains the same functional form as that of a single element,
\begin{equation}
    S_{\mathcal{G}} = \prod_{l=1}^m S(\sigma_l) = \prod_{l=1}^m e^{-\left(\frac{\sigma_l}{\sigma_0}\right)^n} = e^{-\left(\sum_{l=1}^m \sigma_l^n\right)/\sigma_0^n}.
\end{equation}
That is, we can consider the set as an element itself, enforcing that it survives if, and only if, all of its constituents have not failed yet. A schematic of this construction can be seen in Fig.~\ref{fig:Figure3}. This naturally gives rise to a re-normalized stress, as the $n^{\text{th}}$ root of the $L^n$-norm of the vector of stresses $(\sigma_1, \ldots, \sigma_m)$, known in the literature \cite{AndrieuA.2012BmMa} as the \textit{Weibull stress}:
\begin{equation}\label{eq:weibull_stress}
    \bar \sigma_{\mathrm{\mathcal{G}}} = \left( \sum_{l=1}^m \sigma_l^n \right)^{1/n},
\end{equation}
which has a direct analog in the continuum that involves an integral in place of the sum. In the language of the renormalization group \cite{WilsonKennethG.1975TrgC}, this can be interpreted as a change in the effective stress of the system when the scale of the problem is changed---zooming out leads to a lower survival probability, that is nonetheless still captured by Weibull statistics.

For our study, this is important as it means that Eq. (\ref{eq:p_anomalous}) is also relevant for describing the failure statistics of two groups ($\mathcal{G}_1$, $\mathcal{G}_2$) of elements, with $\kappa_{12} = \bar\sigma_{\mathrm{ \mathcal{G}_1}}/\bar\sigma_{\mathrm{\mathcal{G}_2}}$ becoming the ratio of their Weibull stresses. We now elaborate on the utility of this point by considering the probability of diffuse failure.

\subsection{Diffuse Failure}\label{sec:Diffuse_failure}

In particular, we are going to study the probability of another beam anywhere other than the crack tip, failing before the crack tip advances. We consider the set of all lattice elements $\mathcal{L}$ as well as a subset containing all crack-tip elements $\mathcal{C}$. We can call the set of elements in the lattice bulk anywhere other than the crack-tip $\mathcal{B} = \mathcal{L} \setminus \mathcal{C}$, where $\setminus$ represents the set difference operator. Regard a small volume $V_0$ in the bulk containing $m$ elements, that can tile the plane. From Eq. (\ref{eq:weibull_stress}) the probability that this volume $V_0$ survives intact is
\begin{equation}
    S_{V_0} = e^{-\bar \sigma^n/\sigma_0^n},
\end{equation}
where $\bar{\sigma}^n = \sum_{l=1}^m \sigma_l^n$. Assume for a moment that $\bar{\sigma}$ is representative of the whole bulk \footnote{A renormalization can also be performed considering a different $\bar{\sigma}$ at each location ($\exp\left[-(1/V_0) \int_V \left(\bar\sigma(r)/\sigma_0\right)^n \mathrm{d}^3r\right]$), which would just lead to a different definition of $\bar\sigma_{\mathcal{B}}$, but the same qualitative result. We avoid it to de-clutter the notation}. The probability that the bulk set $\mathcal{B}$ (excluding the crack-tip) survives is then
\begin{align}
    S_{\mathcal{B}}(\bar\sigma) = \left(S_{V_0}(\bar{\sigma})\right)^{V/V_0} &= e^{-\frac{V}{V_0}\left(\frac{\bar\sigma}{\sigma_0}\right)^n} \nonumber\\ 
    &= e^{-\left(\frac{(V/V_0)^{1/n} \ \bar\sigma}{\sigma_0}\right)^n} \nonumber\\
    &\equiv e^{-\left(\frac{\bar\sigma_{\mathcal{B}}}{\sigma_0}\right)^n},
    \label{eq:renorm_volume}
\end{align}
where $\bar\sigma_{\mathcal{B}} \equiv (V/V_0)^{1/n} \bar\sigma$, and $V$ is the total bulk volume. As we already discussed, increases in the volume $V$ can effectively be thought of as a combined increase in the internal stresses and a zoom-out operation. Importantly, the survival probability can be made arbitrarily small, for arbitrarily large $V$, if $\bar \sigma>0$. 

In exactly the same way, the survival probability of the crack tip is
\begin{align}
    S_{\mathcal{C}}(\{\sigma_{i \in C} \}) = \prod_{i=1}^6 S(\sigma_i) &= \prod_{i=1}^6 e^{-\left(\frac{\sigma_i}{\sigma_0}\right)^n}\nonumber\\
    &= e^{-\frac{\sigma_1^n}{\sigma_0^n} \left(\sum_{i=1}^6 \kappa_i^n\right)}\nonumber\\
    &\equiv e^{-\left(\frac{\bar\sigma_{\mathcal{C}}}{\sigma_0}\right)^n}.
\end{align}
Now, with reference to Eq. (\ref{eq:p_anomalous}) \& (\ref{eq:weibull_stress}), i.e., treating the bulk as an element at Weibull stress $\bar\sigma_{\mathcal{B}} = (V/V_0)^{1/n} \bar\sigma$ and the crack-tip as another element at Weibull stress $\bar \sigma_{\mathcal{C}} = \sigma_1 \left(\sum_i \kappa_i^n\right)^{1/n}$, we can write down the probability of some element in the bulk $\mathcal{B}$ breaking, before any of the elements in the crack-tip set $\mathcal{C}$ (diffuse failure):
\begin{equation} \label{eq:non_adj_frac}
    P(s_{\mathcal{B}} > s_{\mathcal{C}}) \equiv P^{(0)}_d = \left(\frac{\bar\kappa^n}{1+\bar\kappa^n}\right),
\end{equation}
where
\begin{equation}
    \bar\kappa = \frac{\bar\sigma_{\mathcal{B}}}{\bar \sigma_{\mathcal{C}}} = \frac{ \bar\sigma \ (V/V_0)^{1/n}}{\sigma_1 \left(\sum_i \kappa_i^n\right)^{1/n}}. \label{eq:weibull_bulk_ratio}
\end{equation}

This result condenses everything about the statistics of this problem on two variables: 
\begin{enumerate}
    \item The ratio of the (mean) stress in the bulk, to the peak stress on the crack tip - $\bar\sigma/\sigma_1$. This is exactly proportional to the inverse of a non-dimensionalized \textit{Stress Intensity Factor} (SIF) $Y_{SIF}=K_{\mathrm{SIF}}/\bar\sigma \propto \sigma_1/\bar \sigma$, often called the dimensionless or geometrical SIF \cite{andrasic1984dimensionless}.
    \item The relative stresses of the elements on the crack tip - $\left(\sum_i \kappa_i^n\right)^{1/n} \in [1,6], \ \forall \ n\geq1$. This term becomes larger, the more even the stresses are on the crack-tip, but is still relatively close to 1 for large $n$ (e.g., for $n=4$, the largest it can become is $6^{1/4} \sim 1.565\dots$).
\end{enumerate}

Ignoring the dependence on the latter term, which for our setup ($n \geq 4$) and the \textit{crack-tip} stresses we observe is $ 1 \leq \left(\sum_i \kappa_i^n\right)^{1/n} < 1.16...$, the weibull stress ratio of Eq. (\ref{eq:weibull_bulk_ratio}) takes a form that scales with the SIF and bulk volume \textit{only}:
\begin{equation}\label{eq:k_diffuse}
    \bar \kappa \propto \frac{V^{1/n}}{Y_{\mathrm{SIF}}}.
\end{equation}
Here we see once again that if the volume is large enough, diffuse failure will almost always be observed \textit{before} the crack-tip advances, thus introducing further defects in the bulk that the crack may interact with. Note that in our numerical simulations the specimen volume $V$ is fixed and finite, so $\smash{P^{(0)}_d}$ remains in an intermediate range and does not saturate to 1. Note also that Eq. (\ref{eq:non_adj_frac}) is only exact at the first failure event; as damage nucleates the stress concentrations in the bulk and crack-tip change, leading to a deviation of the statistics from this prediction. Complementary to this analysis, in Appendix \ref{sec:appendix_Evolution_Diffuse}, we motivate and derive the following approximation for the evolution of the diffuse failure probability as the main crack grows:
\begin{equation}\label{eq:diffuse_failure_evolution}
    P_d(N_f) = \frac{P_d^{(0)}}{1 + \beta \ N_f^{\gamma/n}},
\end{equation}
where $\beta$ and $\gamma$ are additional free parameters, and $N_f$ is the number of failed nodes on the main crack. This is a relation we further validate and use in sec. \ref{sec:results}, in order to extract accurate values for $\smash{P^{(0)}_d}$ at the onset of failure.

\section{Mechanically Informed Statistical Modeling}\label{sec:mechanicall_statistical}

\begin{figure*}[ht]
    \centering
    \includegraphics[width=0.9\textwidth]{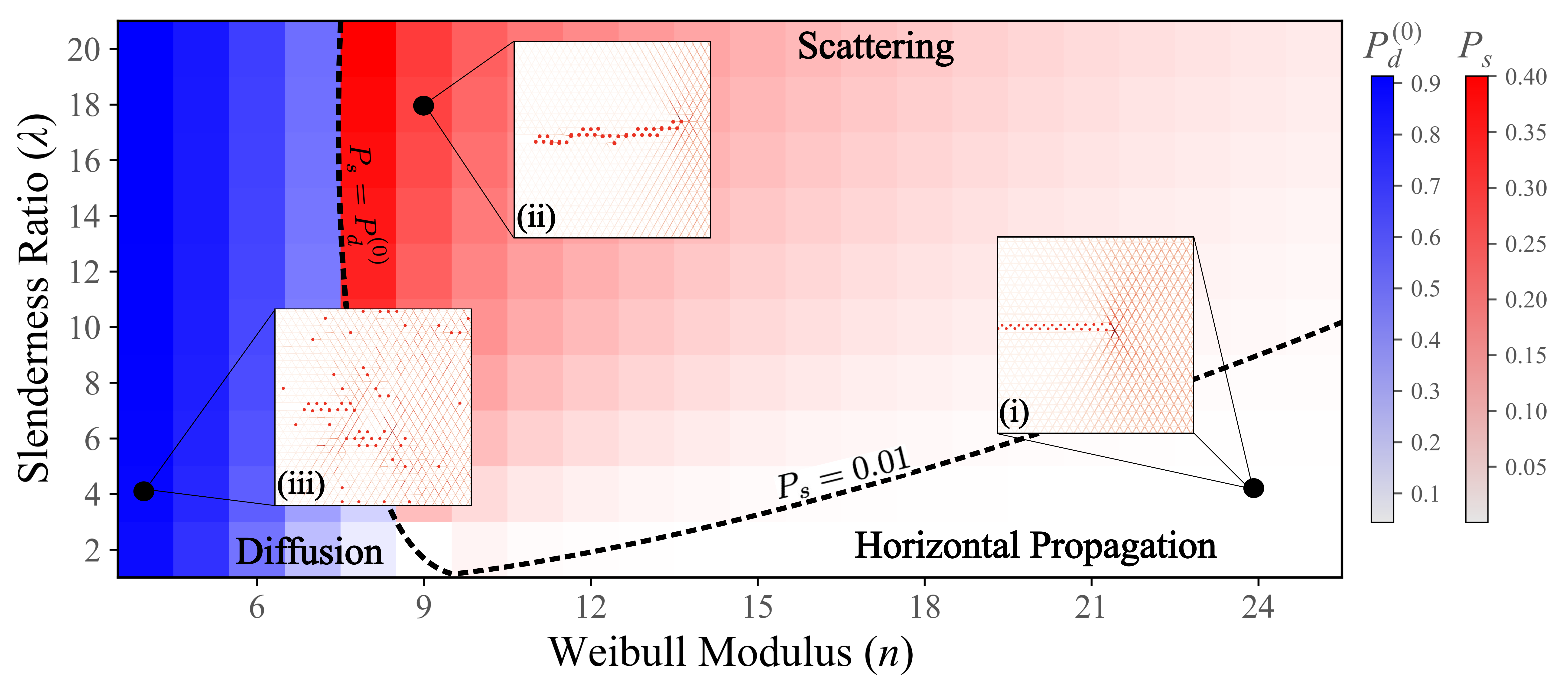} 
    \caption{\justifying Fracture regimes resulting from the mechanically informed statistical analysis. At each point in parameter space, the most probable mode of non-horizontal crack propagation is used to color the corresponding square according to eqs. (\ref{eq:non_adj_frac}) \& (\ref{eq:p_scatter}), with the opacity indicating the associated probability. This construction naturally gives rise to three distinct regions, which are labeled accordingly. Insets show representative failure patterns drawn from each region. Dashed lines are shown as guides to the eye, and delineate the different regimes identified; the conditions defining these boundaries are indicated in text over the dashed lines. At fixed disorder strength, varying the SR alone is sufficient to drive transitions between these regimes, highlighting the possibility for controlling crack-path morphology through geometry.}
    \label{fig:Figure4}
\end{figure*}

We now combine the preceding statistical analysis, i.e., our expressions for $P_s$ and $\smash{P^{(0)}_d}$ (Eqs.~(\ref{eq:p_anomalous}) and (\ref{eq:non_adj_frac})–(\ref{eq:weibull_bulk_ratio})), with the crack-tip stresses $\sigma_{k}(\lambda)$ obtained from an analysis of the crack-tip micromechanics. This allows us to obtain theoretical estimates for the number of failed bonds $N_f$ and for the probability of diffuse failure at the onset of crack propagation $\smash{P_d^{(0)}}$, as functions of $\lambda$ and $n$.

We begin with an estimate of the probability of crack scattering, given that failure has nucleated on one of the crack-tip elements. We introduce a simple binary vector to segregate the \textit{crack-tip} nodes into two sets: the ones whose failure leads to scattering, and the ones that do not. We form this vector with 0s for elements 1 \& 2 and 1s elsewhere, given that scattering is only produced when elements 3-6 fail: $\boldsymbol{h}=[0,0,1,1,1,1]$. Moreover, arranging the six crack-tip stresses into a conjugate vector, raised to the $n^{th}$ \textit{Hadamard} power, i.e., $\boldsymbol{\sigma_{\mathcal{C}}}^{\circ n} (\lambda) = [\sigma^n_1(\lambda), \cdots, \sigma^n_6(\lambda)]$, the probability of scattering ($P_s$) can be neatly written using Eq. (\ref{eq:p_anomalous}) \& (\ref{eq:weibull_stress}):
\begin{equation} \label{eq:p_scatter}
    P_s(n,\lambda) = \frac{\kappa_s^n(\lambda)}{1+\kappa_s^n(\lambda)},
\end{equation}
where
\begin{equation}
    \kappa_s^n(\lambda) = \frac{\boldsymbol{h} \cdot \boldsymbol{\sigma_{\mathcal{C}}}^{\circ n}(\lambda)}{(\mathbf{1}-\boldsymbol{h}) \cdot \boldsymbol{\sigma_{\mathcal{C}}}^{\circ n}(\lambda)}.
\end{equation}

The probability of diffuse failure Eq. (\ref{eq:non_adj_frac}) on the other hand, is expected to have very weak dependence on the SR for our problem setup. The reason is that, as already discussed in sec.\ref{sec:Problem formulation}, we maintain a constant total beam area $A = t \cdot b$, which ensures a constant lattice specific density ($\rho$) across all SR. Therefore, and given that the dominant contributing terms in Eq. (\ref{eq:k_diffuse}) for the stress intensity factor $Y_{\mathrm{SIF}}$ depend only on $\rho$ and not SR \cite{fleck2007damage}, we expect that the dominant contribution to the probability of diffuse failure will also not depend on SR. Nonetheless, we do expect a sub-dominant dependence on SR, e.g., via bending derived micropolar effects \cite{berkache2022micropolar}.

\subsection{Distinct Failure Regimes}

From a study of the probabilities of scattering $P_s$ and diffuse failure $\smash{P^{(0)}_d}$ in the space of $\lambda$ and $n$, we identify distinct regimes for the damage behavior. It is therefore instructive to visualize these on a single diagram. Although the delineation of their boundaries may be arbitrary, given the smooth transition from one regime to another, segregating these regions is still useful and relevant. The difficulty of rigorously differentiating fracture regimes is a problem that has been discussed in existing literature \cite{shekhawat2013damage}, in the context of creating similar phase diagrams.

For our setup, and with reference to Fig.~\ref{fig:Figure4}, the regimes we observe are the following. (i): When $P_s \approx P^{(0)}_d \approx 0$, damage is expected to nucleate at the crack-tip from the beginning and propagate in a horizontal line (no scattering). (ii): For $P_s \gg P^{(0)}_d \approx 0$, damage is expected to nucleate predominantly at the crack-tip, scattering as it advances. With reference to Fig.~\ref{fig:Figure4}, this region is defined by $P_s > P^{(0)}_d$ when $P_s > 0.01$. (iii): For $P^{(0)}_d > P_s$, diffuse failure initially dominates the response.

As expected, we observe that the probability of diffuse failure has almost no dependence on SR (this is also most easily seen in Fig.~\ref{fig:Figure5}(b)). Again, we interpret this as a consequence of the way we have chosen to traverse SR, which fixes the lattice specific density, and thus the SIF in Eq. (\ref{eq:k_diffuse}). On the other hand, the scattering probability strongly depends on SR, and decays much slower as a function of $n$ compared to $P_d$, which is the reason behind the existence of region (ii). This implies that, at least within this region, there is a possibility to control the expression of quenched disorder during failure through one geometric parameter, with larger SR producing higher scattering rates and thus more tortuous crack paths.


In the following section, we use lattice fracture simulations to test the validity of the statistical framework we have presented above for predicting the damage morphology, while also discussing the work of fracture in conjunction.

\section{Results}\label{sec:results}

\subsection{Damage Evolution and Morphology}

We perform simulations of rectangular lattice domains with $N_x = 80$ and $N_y = 40$ unit cells in the x and y directions respectively. Data is collected from $50$ simulations for each pair of $\lambda = \{2,4,7,10,15,20\}$ and $4 \leq n \leq 20, \ n \in \mathbb{Z}$. Each simulation uses a different random seed, so the underlying failure stress distribution is distinct for every run. The lattices are then quasi-statically strained perpendicular to the main crack face, as discussed in Fig.~\ref{fig:Figure1}(a).

\begin{figure*}[htbp]
    \centering
    \includegraphics[width=1.0\textwidth]{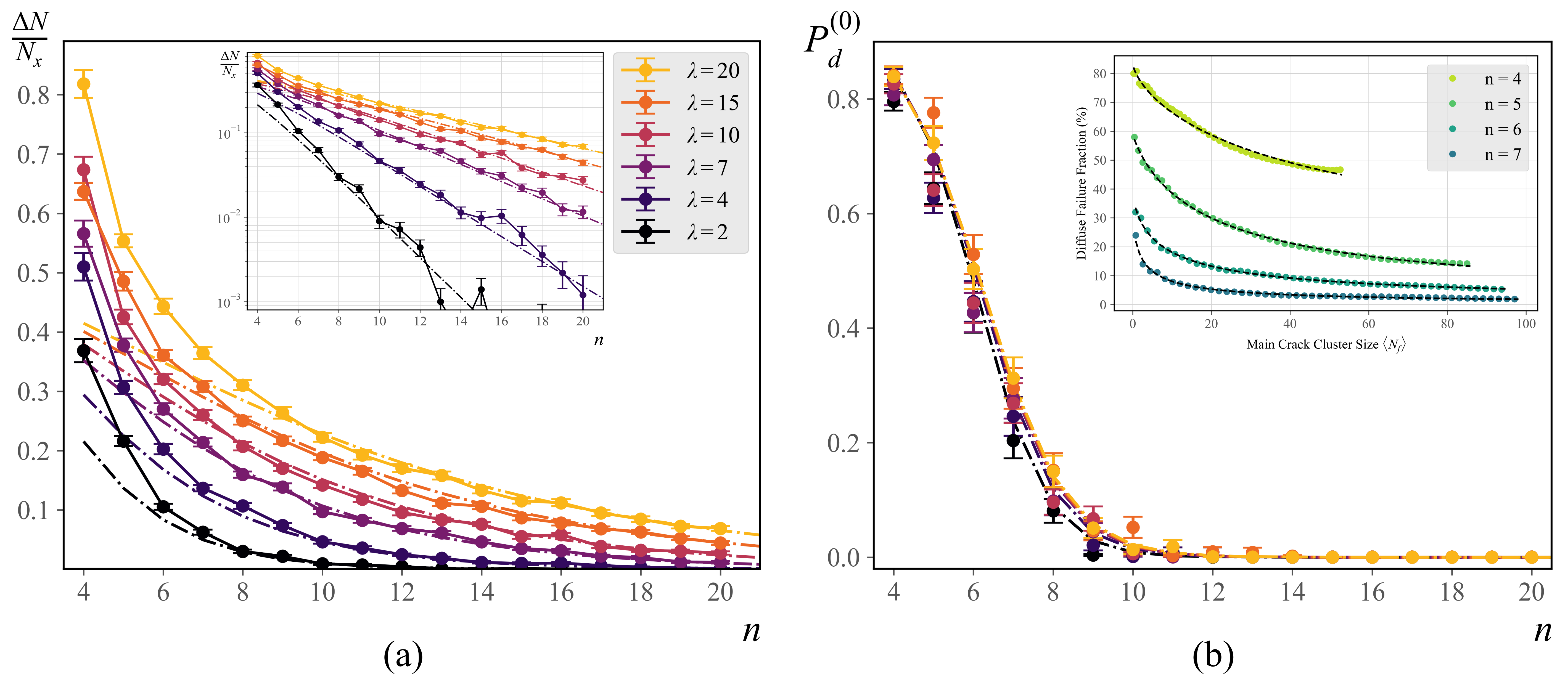} 
    \caption{\justifying Simulation data plotted over theoretical predictions for the probabilities of scattering and diffuse failure. Note that the theoretical (dashed) lines are slices of the respective probabilities from Fig.~\ref{fig:Figure4}, and they have no free parameters; i.e., they are overlaid, not fitted to the data. Error bars indicate the standard error of the mean. (a): Ratio of excess broken bonds $\Delta N = N_f - N_x$ to $N_x$, over the theoretical predictions for $P_s$. In the inset we show the same data in semi-log scale, which clearly demonstrates not only convergence, but also the exponential form for the decay of probability $P_s$ at large $n$ (b): Probability of diffuse failure at the beginning of the damage process---$\smash{P^{(0)}_d}$---, with theoretical prediction overlaid. The weak dependence on SR of this probability is obvious both for theory and simulation data. The inset shows the mean fraction of diffuse failure points as a function of the mean size of the main cluster $N_f$, for $\lambda = 10$, with the dashed lines being fits of Eq. (\ref{eq:diffuse_failure_evolution}); $\smash{P^{(0)}_d}$ in the main plot is extracted as the y-intercept of these curves.
}
    \label{fig:Figure5}
\end{figure*}

Ideally, we expect that the statistical analysis can be related both to the crack path length Eq. (\ref{eq:crackpath_length}) and number of broken bonds on the main crack cluster Eq. (\ref{eq:crack_bonds}). However, the latter is a more robust metric when it comes to extracting it from simulation data. Furthermore, there are challenges when it comes to rigorously defining the notion of crack path length in a discrete system such as a lattice.

Considering the aforementioned difficulties, and the relevance of the number of broken bonds belonging to the main crack, we choose to visualize the fraction of \textit{excess broken bonds} $\Delta N = N_f - N_x$ to $N_x$, for the main crack cluster. This main cluster is isolated from the rest of the damaged bonds using \textit{Hierarchical Agglomerative Clustering} with a Euclidean distance metric \cite{murtagh2012algorithms} (for further details see Appendix \ref{sec:appendix_dataanalysis}). A visualization of the simulation data, together with the theoretically expected scattering probability from Eq. (\ref{eq:p_scatter}) can be seen in Fig.~\ref{fig:Figure5}(a).

We also extract estimates for the probability of diffuse failure at the beginning of the crack propagation $\smash{P^{(0)}_d}$. These are calculated as the y-intercept when fitting Eq. (\ref{eq:diffuse_failure_evolution}) to the average number of diffuse failure events, as a function of the main crack cluster size $N_f$ (an example of fitting these curves can be seen in the inset of Fig.~\ref{fig:Figure5}(b)). We have chosen to extract $\smash{P^{(0)}_d}$ from a fit in order to use a larger amount of data deeper in the failure process, thus obtaining better numerical estimates. Plotted in Fig.~\ref{fig:Figure5}(b), we observe the expected dependence of this probability both on $n$ and SR.

The inset of Fig.~\ref{fig:Figure5}(a) shows that the simulation data closely follow the theoretically predicted scattering probability $P_s$ and converge asymptotically with increasing $n$, while exhibiting clear deviations below a threshold disorder strength. The excellent agreement observed at weak to moderate disorder indicates that successive failure events can be treated as statistically independent, such that crack-path statistics are effectively governed by the statistics of individual failure events. The breakdown of this agreement at low $n$ is thus a notable feature. As seen in Fig.~\ref{fig:Figure5}(b), the onset of the deviation occurs consistently for Weibull moduli in the range $n \simeq 7$–$9$ across all SRs, coinciding with the regime where diffuse failure becomes significant. We speculate that, within this regime the main crack propagates through an already damaged lattice, intermittently linking pre-existing diffuse clusters; scattering events may then be enhanced by the attraction of the crack toward nearby defects, leading to an effectively higher scattering probability than predicted by the independent-event approximation. Despite this deviation, the central role of SR in controlling the mechanical expression of disorder remains clearly evident across the entire range of $n$ explored.


\subsection{Energy Release Rate}

\begin{figure}[ht]
    \centering
    \includegraphics[width=1.0\columnwidth]{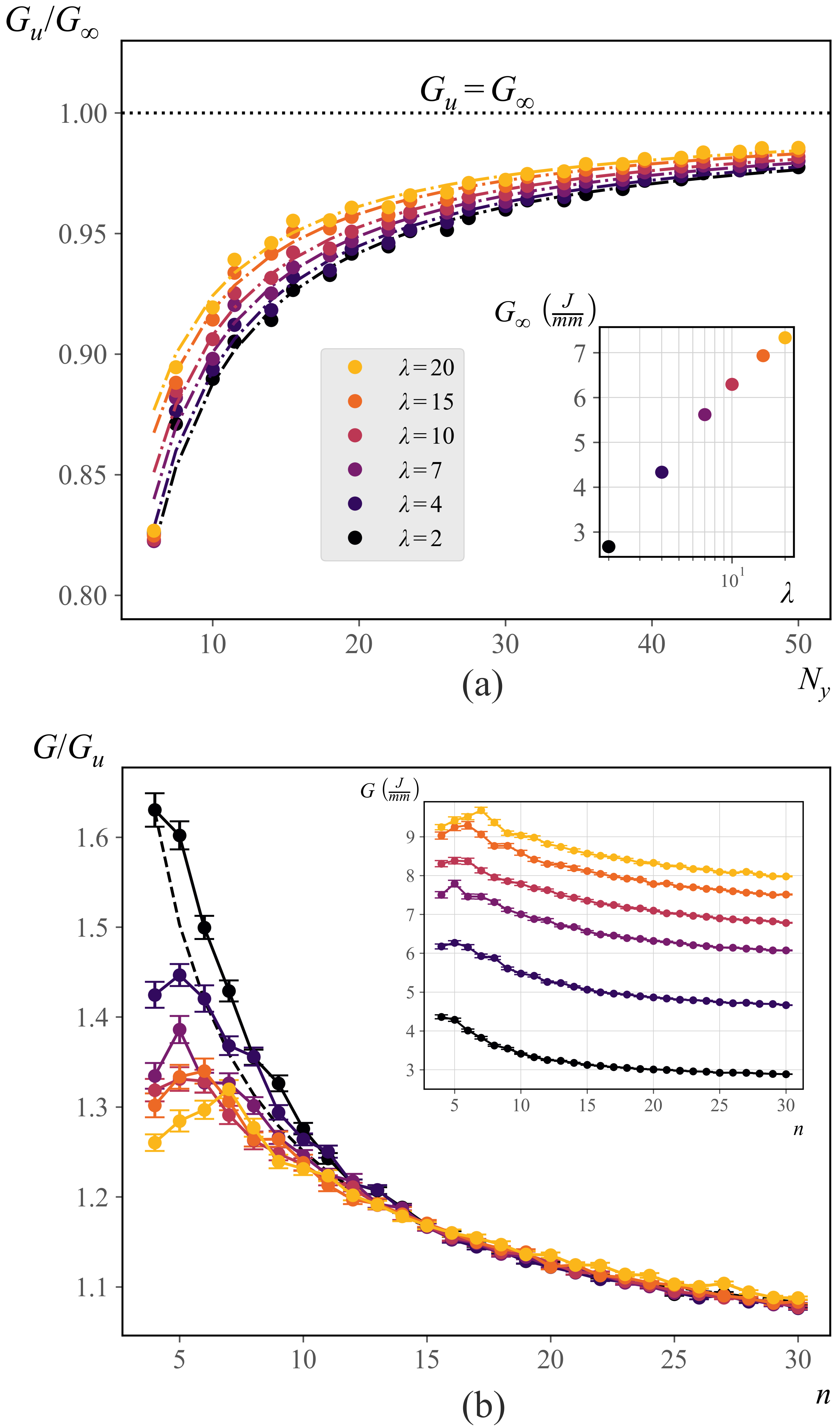} 
    \caption{\justifying Energy release rate $G$ in uniform and disordered lattices. (a): Convergence of $G_u/G_\infty$ with horizontal system size $N_x$ for different SR in uniform---$n\rightarrow\infty$---lattices; the inset shows $G_\infty$ vs SR. The smallest $N_x$ for which $G_u \ge 0.95 G_\infty$ is used in simulations for subsequent measurements in disordered lattices. (b): Normalized fracture energy $G/G_u$ as a function of Weibull modulus $n$ for all SR. Curves collapse for $n \gtrsim 10$, while the inset shows un-collapsed $G$ vs $n$, revealing a non-monotonic dependence on disorder. The collapsed regime is well described by $G/G_u = 1 + B/n$, with the fitted curve plotted as a dashed black line.}
    \label{fig:Figure6}
\end{figure}

We also extract the energy release rate from the simulation data. This quantity characterizes the average energy required to advance the main crack by a distance $s$ and is conventionally defined as $G = \mathrm{d}U/\mathrm{d}s$. To obtain a robust estimate in the presence of crack tortuosity---where bonds between the two crack faces may remain intact until late stages of damage---we compute $G$ as the total fracture energy of a specimen divided by the horizontal distance traversed by the crack. Specifically, this corresponds to the energy required to quasi-statically load the lattice up to the point at which a continuous main crack fully separates the system into two disconnected domains. The choice of the horizontal crack extension is motivated by the predominant mode-I loading of the system, for which, in the absence of disorder, crack propagation is expected to occur along a straight horizontal path.

Finite-size effects play an important role in lattice fracture, making it essential to establish how the above definition of the energy release rate converges as the system size is increased. We therefore study the size dependence of $G$ in the limit $n \rightarrow \infty$, corresponding to a uniform lattice, for which the fracture process is deterministic and yields a well-defined reference value $G_u$. In contrast to the preceding results---where larger lattices could be efficiently simulated to characterize scattering statistics---the accurate evaluation of the work of fracture requires sufficiently fine strain stepping to resolve energy changes, which substantially increases the computational cost. For this reason, the data presented here are drawn from smaller systems, whose dimensions we choose based on a convergence analysis outlined below. Note that, up to a scale factor, the remaining system parameters (e.g. the relative ratio $N_x = 2N_y$, the pre-crack placement along the mid-plane and the loading/boundary conditions) are kept fixed.

For all SR considered, we find that $G_u$ approaches its asymptotic value $G_\infty$ with increasing system size $N_x \times N_y$, and that this convergence is accurately described by the empirical form
\begin{equation}
    G_u = G_{\infty} A^{1/N_y},
\end{equation}
extracted from \cite{li2023specimen}, where $A$ and $G_{\infty}$ are free parameters in the model. Using this relation, we extract $G_\infty$ for each SR (shown in the inset of Fig.~\ref{fig:Figure6}(a)) and verify convergence by plotting $G_u/G_\infty$ as a function of $N_y$ in Fig.~\ref{fig:Figure6}(a). We then select the smallest system size $N_x$ for all subsequent data, for which $G_u$ reaches at least $95\%$ of $G_\infty$ for all SR, balancing numerical accuracy against computational cost; all subsequent measurements of $G$ in disordered lattices are thus collected at a system size of $N_x = 60$, $N_y = 30$, with $100$ runs for each pair of $\lambda = \{2,4,7,10,15,20\}$ and $4 \leq n \leq 20, \ n \in \mathbb{Z}$.

In Fig.~\ref{fig:Figure6}(b), we show the normalized energy release rate $G/G_u$ as a function of the Weibull modulus $n$, for all SR considered. Strikingly, for $n \gtrsim 10$ the data collapse almost perfectly onto a single master curve, indicating that in this regime the disorder-induced modification of the fracture energy becomes effectively independent of SR when measured relative to the uniform reference. The inset of Fig.~\ref{fig:Figure6}(b) presents the corresponding unnormalized values of $G$ as a function of $n$, where the separation between curves makes it evident that, for all $\lambda \geq 4$, the fracture energy exhibits a non-monotonic dependence on disorder strength, attaining a maximum at a relatively low, SR-dependent value of $n$. At present, we do not propose a mechanistic model that captures this non-monotonicity; however, we note that similar trends have been reported in both experimental \cite{fulco2025disorder} and numerical \cite{fulco2025disorder, hartquist2025fracture} studies of fracture with different forms of disorder, suggesting that this behavior may be a robust feature of such systems.

Within the collapsed regime ($n \gtrsim 10$), the data are excellently described by the form represented by the dashed curve in Fig.~\ref{fig:Figure6}(b),
\begin{equation}\label{eq:master_curve_g}
    \frac{G}{G_u} = 1 + \frac{B}{n},
\end{equation}
where $B$ is the single fitted constant. This scaling is consistent with theoretical estimates in which apparent toughening arises from crack arrest induced by spatial variability in local fracture energy \cite{sanner2025less, curtin1990microcrack, charles2002crack}. In such descriptions, crack propagation is assumed to arrest at the point of largest local toughness encountered along its path, leading to an apparent fracture energy that scales with the standard deviation of the underlying toughness distribution. Although we do not explicitly compute fluctuations in local fracture toughness, in the non-diffuse regime crack propagation is governed by the failure of the most critical bond, so that the relevant fluctuations are expected to scale with those of the failure stresses. Accordingly, in the present model, the standard deviation of the Weibull-distributed failure stresses is well approximated by $\sim 1/n$ for $n \geq 4$, providing a natural explanation for the observed scaling of the master curve Eq. (\ref{eq:master_curve_g}). While this description does not account for the non-monotonic behavior observed at lower $n$, it accurately captures the disorder-induced enhancement of fracture energy in the weak-to-moderate disorder regime. Given that within this range we do not expect any initially diffuse failure (Fig.~\ref{fig:Figure5}(b)), we can further support that crack arrest events dominate the energetics of propagation. 


Taken together, these results establish a clear separation between the geometric and statistical contributions to fracture energetics in disordered lattices. While the uniform reference energy $G_u$ is fully determined by lattice geometry through SR, the presence of disorder leads to an apparent enhancement of the fracture energy that collapses onto a universal scaling for sufficiently weak disorder. This feature as well as the emergence of a non-monotonic dependence of $G$ on disorder strength---observed across the SR range---indicates that this enhancement cannot be understood as a simple monotonic consequence of increasing crack tortuosity, or the amount of damaged bonds. In the following discussion, we interpret these findings in the context of crack arrest, damage localization, and the role of geometry in regulating the macroscopic expression of quenched disorder, and outline the broader implications for the design of architected materials.



\section{Final Remarks}
We have combined a statistically informed fracture framework with numerical simulations to demonstrate that the \textit{Slenderness Ratio} (SR) acts as a key geometric control parameter governing how quenched disorder manifests in beam-jointed triangular lattices. Within this framework, we identify three distinct failure regimes in the space of disorder and SR, and show how their emergence is rooted in the geometry-dependent micromechanics at the crack tip. These insights connect with previous work on manufacturing-induced disorder, which is typically addressed through prediction and mitigation: either by anticipating where inhomogeneities will arise \cite{sreejith2021thermodynamic, issametova2024determination, taherkhani2022unsupervised} or by minimizing and tolerating their effects once present \cite{maleki2021surface, yan2019effect}. Our work introduces a complementary perspective: rather than eliminating disorder, we show that its mechanical expression can be actively shaped through geometry, producing lattices that are disorder tolerant under damage. We further identify when such control is most effective, noting that diffuse failure at strong disorder may inhibit the geometric influence we theoretically predict. Overall these observations underpin a key implication of our framework: within a well-defined range of disorder and SR, crack-path morphology can be predicted in a system-size–independent manner from local crack-tip statistics, by treating successive failure events as effectively independent.

Moreover, while our results reproduce the apparent increase in fracture energy reported within the literature of disordered lattices \cite{urabe2010fracture, ziemke2024defect, fulco2025disorder, fulco2025fracture, hartquist2025fracture}, they also reveal that this enhancement cannot be straightforwardly attributed to more distributed damage or increased crack tortuosity under the assumption of a fixed energy of fracture per bond, as alluded to in certain studies \cite{urabe2010fracture, fulco2025disorder}. Instead, our results are consistent with interpretations in which apparent toughening arises from statistical effects associated with local variability in fracture resistance, such as crack arrest and delayed crack advance caused by fluctuations in local failure thresholds \cite{sanner2025less, curtin1990microcrack, charles2002crack}. The geometry-independent scaling of the fracture energy with the variance of the failure stress distribution, supports this interpretation and aligns with theoretical treatments that relate macroscopic fracture energy to the variance of local toughness. At stronger disorder however, where diffuse failure becomes prevalent, this framework breaks down, reflecting a qualitative change in how disorder is expressed at the structural scale. Overall, these results outline the need to distinguish between the effects of disorder on damage morphology, and the physical mechanisms that truly govern energy dissipation during fracture.

This separation between damage morphology and the mechanisms governing energy dissipation outlines a paradigm that goes beyond the prediction or mitigation of disorder, and toward its functionalization: one in which the determinism of crack-path selection can be largely controlled through geometry, independently from the apparent toughening induced by statistical variability. Importantly, this control need not rely solely on manufacturing-induced variability; spatial variations in geometric parameters offer a potential route to deliberately introduce or amplify effective disorder in a controlled manner, suggesting promising directions for future work. More broadly, our results reinforce the notion that disorder, whether inherent or engineered, can be elevated from an unavoidable imperfection to an active design variable, opening new pathways for the design of resilient architected materials with tailored fracture responses.

\section*{Data and Code Availability}
The simulation code and datasets supporting the findings of this article are available at \cite{latticeRepo} and \cite{chouzouris_2026_18482638} respectively.

\begin{acknowledgments}
MC, LdW, and MAD thank UKRI for support under the EPSRC Open Fellowship scheme (Project No. EP/W019450/1). AS, AL, and DSK thank the SNSF for financial support under grants (TMSGI2\_211655 and 200021-236406). All the authors thank Raquel Dantas Batista for the insightful discussions.
\end{acknowledgments}

\bibliography{bibliography}

@book{weibull_statistical_1939,
	address = {Stockholm},
	series = {Ingeniörs vetenskapsakademiens handlingar},
	title = {A statistical theory of the strength of materials},
	url = {http://bibpurl.oclc.org/web/34398 http://dds.crl.edu/CRLdelivery.asp?tid=9643},
	language = {english},
	urldate = {2025-05-29},
	publisher = {Generalstabens litografiska anstalts förlag},
	author = {Weibull, Waloddi},
	year = {1939},
	note = {OCLC: 30416455},
}

@article{AndrieuA.2012BmMa,
author = {Andrieu, A. and Pineau, A. and Besson, J. and Ryckelynck, D. and Bouaziz, O.},
address = {OXFORD},
copyright = {2012 Elsevier Ltd},
issn = {0013-7944},
journal = {Engineering fracture mechanics},
language = {english},
pages = {102-117},
publisher = {Elsevier Ltd},
title = {Beremin model: Methodology and application to the prediction of the Euro toughness data set},
volume = {95},
year = {2012},
}

@article{WilsonKennethG.1975TrgC,
author = {Wilson, Kenneth G.},
address = {COLLEGE PK},
copyright = {Copyright 2025 Elsevier B.V., All rights reserved.},
issn = {0034-6861},
journal = {Reviews of modern physics},
language = {english},
number = {4},
pages = {773-840},
publisher = {Amer Physical Soc},
title = {The renormalization group: Critical phenomena and the Kondo problem},
volume = {47},
year = {1975},
}

@article{de2025architecting,
  title={Architecting mechanisms of damage in topological metamaterials},
  author={de Waal, Leo and Chouzouris, Matthaios and Dias, Marcelo A},
  journal={Physical Review Research},
  volume={7},
  number={3},
  pages={033177},
  year={2025},
  publisher={APS}
}

@article{de2025cracking,
  title={Cracking Down on Fracture to Functionalize Damage},
  author={de Waal, Leo and Chouzouris, Matthaios and Dias, Marcelo A},
  journal={Physical Review Letters},
  volume={135},
  number={14},
  pages={148202},
  year={2025},
  publisher={APS}
}

@article{LyuYongtao2024Anmm,
author = {Lyu, Yongtao and Song, Xiaoshuang and Wang, Hao and Jiang, Jian},
copyright = {2024 Elsevier Ltd},
issn = {2352-4928},
journal = {Materials today communications},
language = {english},
pages = {110135-},
publisher = {Elsevier Ltd},
title = {A novel mechanical metamaterial with tunable stiffness and individually adjustable poisson’s ratio},
volume = {40},
year = {2024},
}

@article{ZhengXiaoyu2014UUMM,
author = {Zheng, Xiaoyu and Lee, Howon and Weisgraber, Todd H. and Shusteff, Maxim and DeOtte, Joshua and Duoss, Eric B. and Kuntz, Joshua D. and Biener, Monika M. and Ge, Qi and Jackson, Julie A. and Kucheyev, Sergei O. and Fang, Nicholas X. and Spadaccini, Christopher M.},
address = {United States},
copyright = {Copyright © 2014 American Association for the Advancement of Science},
issn = {0036-8075},
journal = {Science (American Association for the Advancement of Science)},
language = {english},
number = {6190},
pages = {1373-1377},
publisher = {American Association for the Advancement of Science},
title = {Ultralight, Ultrastiff Mechanical Metamaterials},
volume = {344},
year = {2014},
}

@article{SchwaigerR.2019Temo,
author = {Schwaiger, R. and Meza, L.R. and Li, X.},
address = {New York, USA},
copyright = {Copyright © Materials Research Society 2019},
issn = {0883-7694},
journal = {MRS bulletin},
language = {english},
number = {10},
pages = {758-765},
publisher = {Cambridge University Press},
title = {The extreme mechanics of micro- and nanoarchitected materials},
volume = {44},
year = {2019},
}

@article{KhosravaniMohammadReza2025Fomm,
author = {Khosravani, Mohammad Reza and Anders, Denis and Ayatollahi, Majid R. and Reinicke, Tamara},
address = {London},
copyright = {The Author(s) 2025},
issn = {2083-3318},
journal = {Archives of Civil and Mechanical Engineering},
language = {english},
number = {5-6},
pages = {244-},
publisher = {Springer London},
title = {Fabrication of mechanical metamaterials by 3D printing: recent advancements and current challenges},
volume = {25},
year = {2025},
}

@article{ZhangHang2018Smmw,
author = {Zhang, Hang and Guo, Xiaogang and Wu, Jun and Fang, Daining and Zhang, Yihui},
address = {United States},
issn = {2375-2548},
journal = {Science advances},
language = {english},
number = {6},
pages = {eaar8535-},
publisher = {American Association for the Advancement of Science},
title = {Soft mechanical metamaterials with unusual swelling behavior and tunable stress-strain curves},
volume = {4},
year = {2018}
}

@article{RafsanjaniAhmad2016Bamm,
author = {Rafsanjani, Ahmad and Pasini, Damiano},
issn = {2352-4316},
journal = {Extreme Mechanics Letters},
language = {english},
pages = {291-296},
title = {Bistable auxetic mechanical metamaterials inspired by ancient geometric motifs},
volume = {9},
year = {2016},
}

@article{masuo2018influence,
  title={Influence of defects, surface roughness and HIP on the fatigue strength of Ti-6Al-4V manufactured by additive manufacturing},
  author={Masuo, Hiroshige and Tanaka, Yuzo and Morokoshi, Shotaro and Yagura, Hajime and Uchida, Tetsuya and Yamamoto, Yasuhiro and Murakami, Yukitaka},
  journal={International Journal of Fatigue},
  volume={117},
  pages={163--179},
  year={2018},
  publisher={Elsevier}
}

@article{silva2025investigation,
  title={Investigation of Distortion, Porosity and Residual Stresses in Internal Channels Fabricated in Maraging 300 Steel by Laser Powder Bed Fusion},
  author={Silva, Bruno Caetano dos Santos and Callegari, Bruna and Seixas, Lu{\~a} Fonseca and Kr{\'o}l, Mariusz and Sitek, Wojciech and Matula, Grzegorz and Krzemi{\'n}ski, {\L}ukasz and Coelho, Rodrigo Santiago and Batalha, Gilmar Ferreira},
  journal={Materials},
  volume={18},
  number={5},
  pages={1019},
  year={2025},
  publisher={MDPI}
}

@article{lertthanasarn2020mechanical,
  title={Mechanical behaviour of additively manufactured Ti6Al4V meta-crystals containing multi-scale hierarchical lattice structures},
  author={Lertthanasarn, Jedsada and Liu, Chen and Pham, Minh-Son},
  journal={arXiv preprint arXiv:2011.14201},
  year={2020}
}

@article{sreejith2021thermodynamic,
  title={A thermodynamic framework for additive manufacturing, using amorphous polymers, capable of predicting residual stress, warpage and shrinkage},
  author={Sreejith, P and Kannan, K and Rajagopal, KR},
  journal={International Journal of Engineering Science},
  volume={159},
  pages={103412},
  year={2021},
  publisher={Elsevier}
}

@article{issametova2024determination,
  title={Determination of residual stresses in 3D-printed polymer parts},
  author={Issametova, Madina and Martyushev, Nikita V and Zhastalap, Abilkaiyr and Sabirova, Layla B and Assemgul, Uderbayeva and Tursynbayeva, Arailym and Abilezova, Gazel},
  journal={Polymers},
  volume={16},
  number={14},
  pages={2067},
  year={2024}
}

@article{taherkhani2022unsupervised,
  title={An unsupervised machine learning algorithm for in-situ defect-detection in laser powder-bed fusion},
  author={Taherkhani, Katayoon and Eischer, Christopher and Toyserkani, Ehsan},
  journal={Journal of Manufacturing Processes},
  volume={81},
  pages={476--489},
  year={2022},
  publisher={Elsevier}
}

@article{maleki2021surface,
  title={Surface post-treatments for metal additive manufacturing: Progress, challenges, and opportunities},
  author={Maleki, Erfan and Bagherifard, Sara and Bandini, Michele and Guagliano, Mario},
  journal={Additive Manufacturing},
  volume={37},
  pages={101619},
  year={2021},
  publisher={Elsevier}
}

@article{yan2019effect,
  title={Effect of hot isostatic pressing (HIP) treatment on the compressive properties of Ti6Al4V lattice structure fabricated by selective laser melting},
  author={Yan, Xingchen and Lupoi, Rocco and Wu, Hongjian and Ma, Wenyou and Liu, Min and O'Donnell, Garrett and Yin, Shuo},
  journal={Materials Letters},
  volume={255},
  pages={126537},
  year={2019},
  publisher={Elsevier}
}

@article{fulco2025disorder,
  title={Disorder enhances the fracture toughness of 2D mechanical metamaterials},
  author={Fulco, Sage and Budzik, Michal K and Xiao, Hongyi and Durian, Douglas J and Turner, Kevin T},
  journal={PNAS nexus},
  volume={4},
  number={2},
  pages={pgaf023},
  year={2025},
  publisher={Oxford University Press US}
}

@article{ziemke2024defect,
  title={The defect sensitivity of brittle truss-based metamaterials},
  author={Ziemke, Patrick and Finney, Owen and Chambers, Ryan G and Thiraux, Raphael and Valdevit, Lorenzo and Begley, Matthew R},
  journal={Materials \& Design},
  volume={239},
  pages={112776},
  year={2024},
  publisher={Elsevier}
}

@article{fulco2025fracture,
  title={Fracture of disordered and stochastic lattice materials},
  author={Fulco, Sage and Purohit, Prashant K and Budzik, Michal K and Turner, Kevin T},
  journal={arXiv preprint arXiv:2508.21187},
  year={2025}
}

@article{bertalan2014fracture,
  title={Fracture strength: stress concentration, extreme value statistics, and the fate of the Weibull distribution},
  author={Bertalan, Zsolt and Shekhawat, Ashivni and Sethna, James P and Zapperi, Stefano},
  journal={Physical Review Applied},
  volume={2},
  number={3},
  pages={034008},
  year={2014},
  publisher={APS}
}

@article{shekhawat2013damage,
  title={From damage percolation to crack nucleation through finite size criticality},
  author={Shekhawat, Ashivni and Zapperi, Stefano and Sethna, James P},
  journal={Physical review letters},
  volume={110},
  number={18},
  pages={185505},
  year={2013},
  publisher={APS}
}

@article{alava2008role,
  title={Role of disorder in the size scaling of material strength},
  author={Alava, Mikko J and Nukala, Phani KVV and Zapperi, Stefano},
  journal={Physical review letters},
  volume={100},
  number={5},
  pages={055502},
  year={2008},
  publisher={APS}
}

@article{berkache2022micropolar,
  title={Micropolar effects on the effective elastic properties and elastic fracture toughness of planar lattices},
  author={Berkache, Kamel and Phani, Srikantha and Ganghoffer, Jean-Fran{\c{c}}ois},
  journal={European Journal of Mechanics-A/Solids},
  volume={93},
  pages={104489},
  year={2022},
  publisher={Elsevier}
}

@article{guenole2020assessment,
  title={Assessment and optimization of the fast inertial relaxation engine (fire) for energy minimization in atomistic simulations and its implementation in lammps},
  author={Gu{\'e}nol{\'e}, Julien and N{\"o}hring, Wolfram G and Vaid, Aviral and Houll{\'e}, Fr{\'e}d{\'e}ric and Xie, Zhuocheng and Prakash, Aruna and Bitzek, Erik},
  journal={Computational Materials Science},
  volume={175},
  pages={109584},
  year={2020},
  publisher={Elsevier}
}

@article{bitzek2006structural,
  title={Structural relaxation made simple},
  author={Bitzek, Erik and Koskinen, Pekka and G{\"a}hler, Franz and Moseler, Michael and Gumbsch, Peter},
  journal={Physical review letters},
  volume={97},
  number={17},
  pages={170201},
  year={2006},
  publisher={APS}
}

@article{luan2022energy,
  title={Energy-based fracture mechanics of brittle lattice materials},
  author={Luan, Shengzhi and Chen, Enze and Gaitanaros, Stavros},
  journal={Journal of the Mechanics and Physics of Solids},
  volume={169},
  pages={105093},
  year={2022},
  publisher={Elsevier}
}

@article{lingua2025breaking,
  title={Breaking better: Imperfections increase fracture resistance in architected lattices},
  author={Lingua, Alessandra and Sanner, Antoine and Hild, Fran{\c{c}}ois and Kammer, David S},
  journal={arXiv preprint arXiv:2504.08873},
  year={2025}
}

@article{weihull1951statistical,
  title={A statistical distribution function of wide applicability},
  author={Weihull, W},
  journal={J Appl Mech},
  volume={18},
  pages={290--293},
  year={1951}
}

@article{andrasic1984dimensionless,
  title={Dimensionless stress intensity factors for cracked thick cylinders under polynomial crack face loadings},
  author={Andrasic, CP and Parker, AP},
  journal={Engineering Fracture Mechanics},
  volume={19},
  number={1},
  pages={187--193},
  year={1984},
  publisher={Elsevier}
}

@article{murtagh2012algorithms,
  title={Algorithms for hierarchical clustering: an overview},
  author={Murtagh, Fionn and Contreras, Pedro},
  journal={Wiley interdisciplinary reviews: data mining and knowledge discovery},
  volume={2},
  number={1},
  pages={86--97},
  year={2012},
  publisher={Wiley Online Library}
}

@article{urabe2010fracture,
  title={Fracture toughness and maximum stress in a disordered lattice system},
  author={Urabe, Chiyori and Takesue, Shinji},
  journal={Physical Review E—Statistical, Nonlinear, and Soft Matter Physics},
  volume={82},
  number={1},
  pages={016106},
  year={2010},
  publisher={APS}
}

@book{timoshenko2012theory,
  title={Theory of elastic stability},
  author={Timoshenko, Stephen P and Gere, James M},
  year={2012},
  publisher={Courier Corporation}
}

@article{fleck2007damage,
  title={The damage tolerance of elastic--brittle, two-dimensional isotropic lattices},
  author={Fleck, Norman A and Qiu, XinMing},
  journal={Journal of the Mechanics and Physics of Solids},
  volume={55},
  number={3},
  pages={562--588},
  year={2007},
  publisher={Elsevier}
}

@article{li2023specimen,
  title={Specimen size effect on the fracture energy of architected stretchable materials},
  author={Li, Xiao and Men, Libo and Yu, Yilin and Hou, Zhaoyang and Wang, Zhengjin},
  journal={International Journal of Smart and Nano Materials},
  volume={14},
  number={4},
  pages={420--439},
  year={2023},
  publisher={Taylor \& Francis}
}

@article{curtin1990microcrack,
  title={Microcrack toughening?},
  author={Curtin, WA and Futamura, K},
  journal={Acta Metallurgica et Materialia},
  volume={38},
  number={11},
  pages={2051--2058},
  year={1990},
  publisher={Elsevier}
}

@article{sanner2025less,
  title={Less is more: removing a single bond increases the toughness of elastic networks},
  author={Sanner, Antoine and Michel, Luca and Lingua, Alessandra and Kammer, David S},
  journal={International Journal of Fracture},
  volume={249},
  number={4},
  pages={1--12},
  year={2025},
  publisher={Springer}
}

@article{charles2002crack,
  title={On crack arrest in ceramic/metal assemblies},
  author={Charles, Yann and Hild, Fran{\c{c}}ois},
  journal={International journal of fracture},
  volume={115},
  number={3},
  pages={251--272},
  year={2002},
  publisher={Springer}
}

@article{hartquist2025fracture,
  title={Fracture of polymer-like networks with hybrid bond strengths},
  author={Hartquist, Chase M and Wang, Shu and Deng, Bolei and Beech, Haley K and Craig, Stephen L and Olsen, Bradley D and Rubinstein, Michael and Zhao, Xuanhe},
  journal={Journal of the Mechanics and Physics of Solids},
  volume={195},
  pages={105931},
  year={2025},
  publisher={Elsevier}
}

@misc{latticeRepo,
  author       = {Matthaios Chouzouris},
  title        = {SimLaD---Simulator for Lattice Damage},
  year         = {2026},
  publisher    = {GitHub},
  howpublished = {\url{https://github.com/McHouzou/SimLaD.git}},
}

@dataset{chouzouris_2026_18482638,
  author    = {Chouzouris, Matthaios and
               de~Waal, Leo and
               Sanner, Antoine and
               Lingua, Alessandra and
               Kammer, David and
               Dias, Marcelo~A.},
  title     = {How Geometry Tames Disorder in Lattice Fracture: Data},
  year      = {2026},
  publisher = {Zenodo},
  doi       = {10.5281/zenodo.18482638},
  url       = {https://doi.org/10.5281/zenodo.18482638}
}

\clearpage
\newpage
\appendix

\section{Method}\label{sec:appendix_method}

\begin{figure}[ht]
    \centering
    \includegraphics[width=0.9\columnwidth]{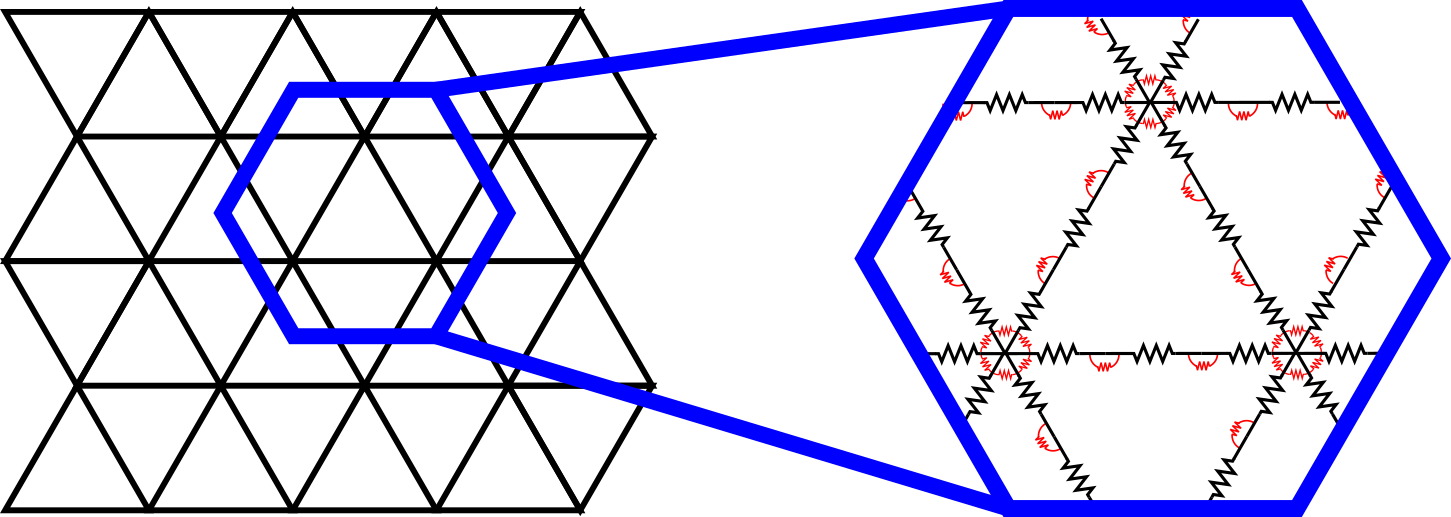} 
    \caption{\justifying Discretization of a triangular lattice, with each beam divided into three independent elements. When disorder is included, all three elements of a beam share the same failure stress.}
    \label{fig:FigureA3}
\end{figure}

We adopt the modeling approach described in~\cite{de2025architecting}, which has been experimentally validated for the fracture of PMMA-based Maxwell networks~\cite{de2025cracking}. The simulator used in this study is publicly available in \cite{latticeRepo}. The system is represented as a mass-spring network with pinned boundaries, incorporating both axial and torsional springs. Each bond is discretized into three elements (see Fig.~\ref{fig:FigureA3}) to capture the dominant low-energy bending modes. Equilibrium at each quasi-static strain step is achieved using the FIRE 2.0 algorithm~\cite{bitzek2006structural,guenole2020assessment}.

Fracture is simulated by splitting nodes when a connected element satisfies the combined axial–bending stress criterion,
\begin{equation}\label{Eq:7} \sigma_\mathrm{F}=\left|\frac{\sigma_\mathrm{A}}{\sigma_\mathrm{A}^\mathrm{u}}\right|+\left|\frac{\sigma_\mathrm{B}}{\sigma_\mathrm{B}^\mathrm{u}}\right| \geq 1, \end{equation} 
where $\sigma_\mathrm{A}$ and $\sigma_\mathrm{B}$ are the axial and bending stresses in the element and $\sigma_\mathrm{A}^\mathrm{u}$ and $\sigma_\mathrm{B}^\mathrm{u}$ represent the element's failure stresses in axial and bending.

\begin{figure}[ht]
    \centering
    \includegraphics[width=1.0\columnwidth]{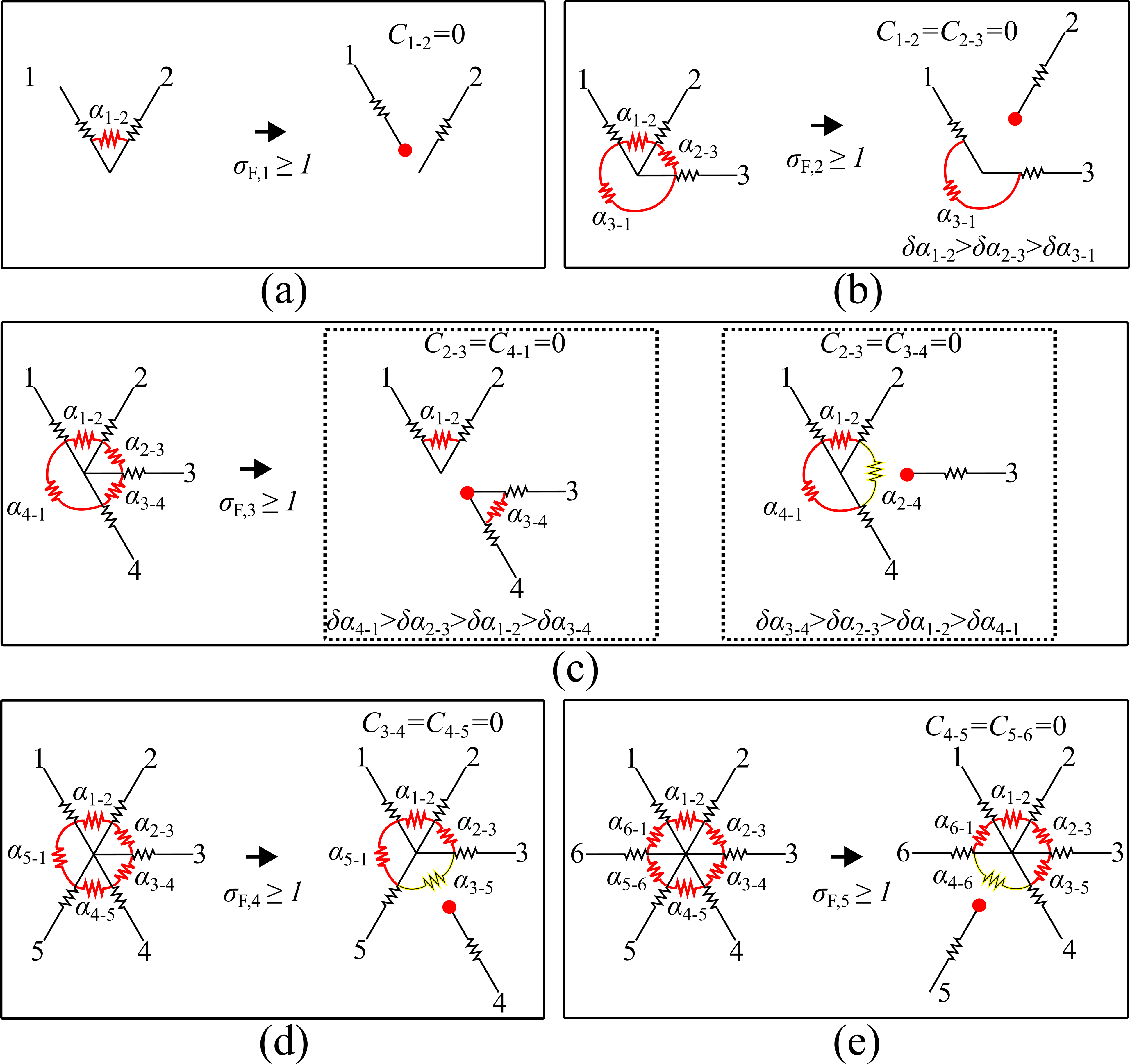} 
    \caption{\justifying Splitting of elements in a lattice when element $i$ reaches failure (i.e., $\sigma_\mathrm{F,i}\geq1$) for nodes with: (a) 2, (b) 3, (c) 4, (d) 5, and (e) 6 elements meeting at a node. Here, $\alpha_\mathrm{i-j}$ and $C_\mathrm{i-j}$ represent the angle and torsional spring stiffness between elements $i$ and $j$, respectively, while $\delta\alpha_\mathrm{i-j}$ denotes the change in angle following loading.} 
    \label{fig:FigureA4}
\end{figure}

Upon node splitting, elements are reassigned to reflect fracture (Fig.~\ref{fig:FigureA4}). The reassignment depends on the number of elements initially connected at the node. For nodes with fewer than five connected elements, the protocol follows that outlined in~\cite{de2025architecting} for critically coordinated lattices. Specifically, for sites with three or four connected elements (Fig.~\ref{fig:FigureA4}(b--c)), we disconnect the elements between the two torsional springs experiencing the largest angular change, keeping elements that rotate together grouped. A new torsional spring is introduced when necessary to maintain local connectivity. For nodes with more than four connected elements (Fig.~\ref{fig:FigureA4}(d--e)), only the single element meeting the failure criterion is separated, and a torsional spring is added to bridge the elements directly beside it. This approach is motivated by experimental observations~\cite{luan2022energy, lingua2025breaking}.

Damage evolves iteratively within each displacement step, where in each iteration, only one site may split, followed by energy minimisation to capture the stress redistribution. If the relaxation immediately after a failure event leads to multiple bonds meeting the failure criteria (i.e., an avalanche of failures), only the most stressed element is split. Otherwise, we use a binary search to reach a strain at which only one element exceeds the failure threshold; the element is subsequently broken. The material is assumed linear-elastic until failure.

\section{Scattering Map} \label{sec:appendix_scattering_map}
As mentioned in the main text, failure of each one of the 6 crack-tip elements either causes the crack to change row, or to continue propagating along the same one. Furthermore, we discussed that failure of any element other than the first two, results in an additional damage point on the main cluster. In other words, if any element other than $1$ or $2$ fails, the main crack has to remove an additional element to reach the same $x$ location as a crack that did not experience such an anomalous failure event. As discussed in the main text this information (of whether scattering or a row change is observed) can be encoded in two matrices containing binary values. We visualize these below in Fig.~\ref{fig:FigureA1}.


\begin{figure}[ht]
    \centering
    \includegraphics[width=0.7\columnwidth]{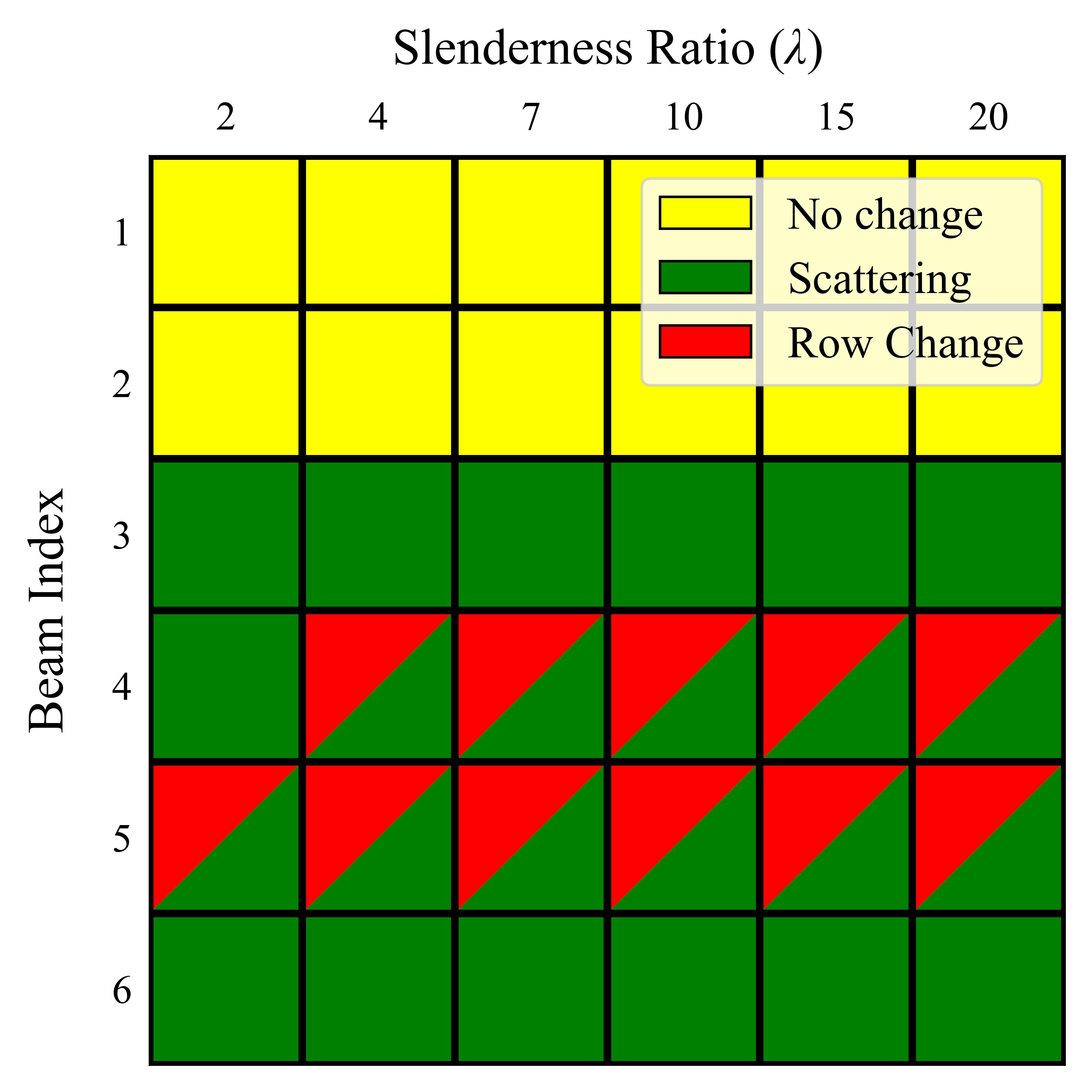} 
    \caption{\justifying Visualization of scattering map and map of row change overlaid, capturing the ensuing damage pattern after the removal of each one of the $6$ crack-tip elements. Row change map specific to PMMA; Scattering map specific to triangular lattice and boundary conditions.}
    \label{fig:FigureA1}
\end{figure}

\section{Damage Morphology from Data} \label{sec:appendix_dataanalysis}
Extracting information about the damage morphology from simulation data is not a straightforward task---certain assumptions and choices are necessary. Here we outline our methodology in detail. 

Before extracting any fracture related metrics from the cloud of damage points, we perform a pre-processing where each one of them is translated back to its respective position at the beginning of the simulation, before the lattice was strained. This is to restore the information about the damage locations in the network, without including details about their relative displacements under the elastic deformation incurred.

\begin{figure}[hb!]
    \centering
    \includegraphics[width=1.0\columnwidth]{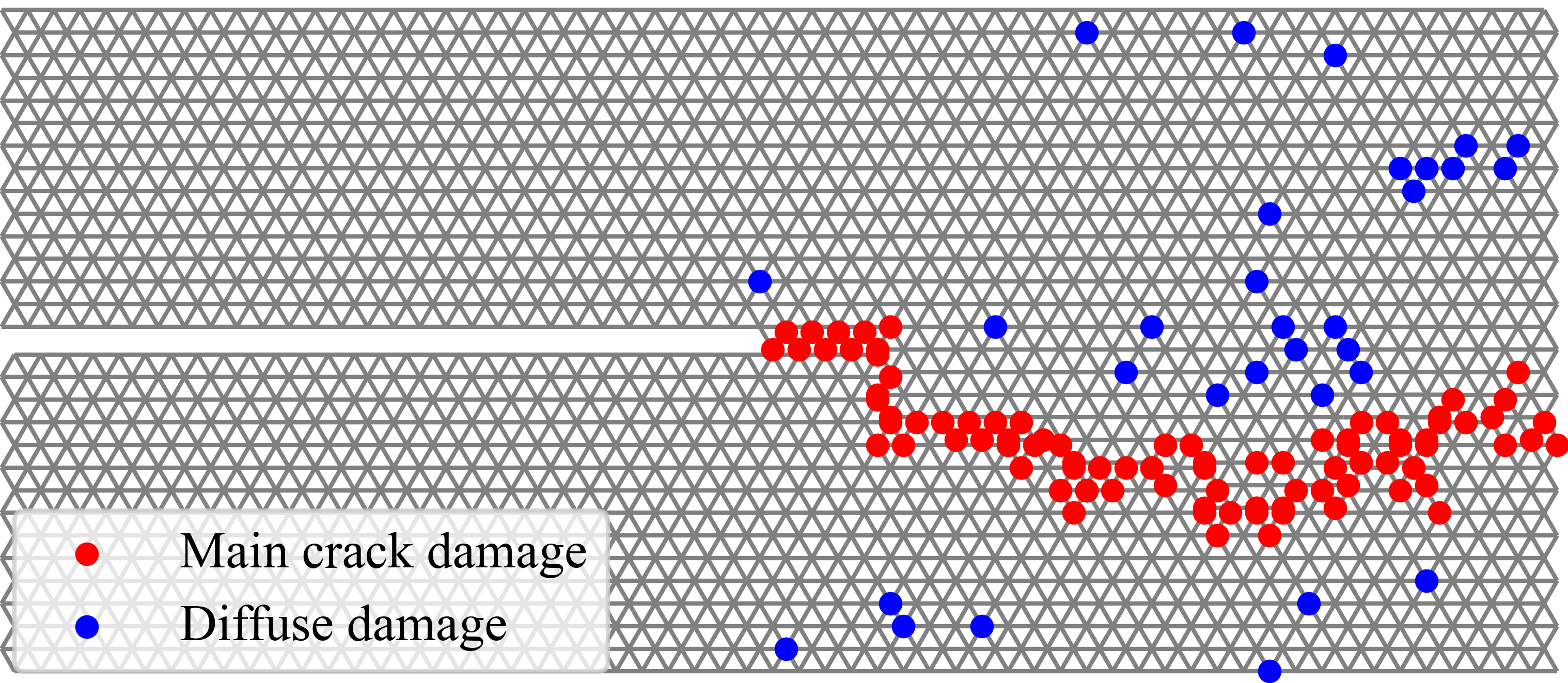} 
    \caption{\justifying Resulting damage point segregation after performing \textit{Hierarchical Agglomerative Clustering}.}
    \label{fig:FigureA2}
\end{figure}

As mentioned in the text, the main crack cluster is isolated from the damage cloud using \textit{Hierarchical Aglomerative Clustering} with a Euclidean distance metric \cite{murtagh2012algorithms}. This algorithm starts by considering all points as separate clusters, iteratively joining the two clusters that are the closest to each other at each step. The algorithm terminates once a certain criterion is reached. In our case, this criterion is that there are no clusters closer than a distance of $1.1 a$ apart, where $a$ is the unit-cell size. This choice was made by noticing that sequential damage events in a horizontally propagating crack, are at most one unit cell apart, and allowing for a small 10$\%$ error so as to avoid accidentally separating clusters that belong to the same group. The number $N_f$ is then simply found by counting the amount of points in the largest cluster. A visualization of the resulting segregation of damage points can be seen in Fig.~\ref{fig:FigureA2}.

\section{Non-linear Corrections} \label{sec:Appendix_nonlinear}
One of the main underlying assumptions in our study has been that the stresses in all beams scale linearly with each-other (that is, we said $\sigma_i = \kappa_{ij} \sigma_j$ with $\kappa_{ij}$ being a constant). Evidently in our chosen system this assumption is justified, but this is not generally true. Here we will examine the implications of relaxing this assumption and will thus supply corrections to the probability Eq. (\ref{eq:p_anomalous}), in the presence of non-linearities.

Let us define a control variable $\delta$, which we can take to represent the fractional $y$ displacement of the top boundary. Then the stress in element $i$, can be written $\sigma_i \equiv \sigma_i(\delta)$. As a result, the stress ratios we defined before will now generally depend on $\delta$. Expanding around $\delta = 0$:
\begin{align}
    \kappa_{ij}(\delta) &= \frac{\sigma_i(\delta)}{\sigma_j(\delta)}\nonumber \\
    &= \frac{\delta \ \sigma_i'(0) + \frac{\delta^2}{2} \sigma_i ''(0) + \mathcal{O}(\delta^3)}{\delta \ \sigma_j'(0) + \frac{\delta^2}{2} \sigma_j ''(0) + \mathcal{O}(\delta^3)} \nonumber \\
    &=\frac{\sigma_i'(0)}{\sigma_j'(0)} + \delta \frac{(\sigma_i''(0) \sigma_j'(0) - \sigma_j''(0) \sigma_i'(0))}{(\sigma_j'(0))^2}+\mathcal{O}(\delta^2) \nonumber \\
    &\equiv \kappa_{ij}^{(0)} + \delta \frac{[\sigma_i''(0) - \kappa_{ij}^{(0)} \sigma_j''(0)]}{\sigma_j'(0)} + \mathcal{O}(\delta^2) .\label{eq:kappa_nonlinear}
\end{align}
Here, $\kappa_{ij}^{(0)}$ is the stress ratio at the limit $\delta \rightarrow 0$. The linear scaling assumption is essentially the same as saying $\sigma_i(\delta) = \sigma^*_i \ f(\delta), \ \forall i$, with $\sigma^*_i$ a constant specific to each element, and $f(\delta)$ a global function. One can check that this retrieves a constant $\kappa_{ij}(\delta) = \sigma_i^*/\sigma_j^*$. In linear elasticity there is the further assumption that $f(\delta) \propto \delta$. So assuming $\kappa_{ij}$ is a constant wrt. the stresses, is more general than assuming a linear elastic material.

To relax the assumption of linear scaling, and also obtain a form for $\kappa_{ij}$ that is workable within the framework laid out up to now, we consider $\sigma_j(\delta)$ being a monotonic increasing function of $\delta$. As a result, we can change variables in (\ref{eq:kappa_nonlinear}) $\delta \rightarrow \sigma_j$, using an inverse function. Employing a series reversion of
\begin{align*}
    \sigma_j(\delta) &= \alpha^{(1)}_j \delta + \alpha^{(2)}_j \delta^2 + ... \\
    \Rightarrow \delta(\sigma_j) &= \frac{1}{\alpha^{(1)}_j} \sigma_j + \left(\frac{\alpha^{(2)}_j}{\left(\alpha^{(1)}_j\right)^3}\right) \sigma_j^2 + ...
\end{align*}

we obtain:
\begin{align}
    \kappa_{ij}(\sigma_j) \approx \kappa_{ij}^{(0)} \ (1 \ + \ \epsilon \ \sigma_j),  \label{eq:expansion_kappa}
\end{align}

where the parameter $\epsilon$ has dimensions of inverse stress---we also consider a non-dimensional $\tilde \epsilon$:
\begin{align}
    \epsilon &= \frac{1}{\left(\sigma_j'(0)\right)^2}\left[ \frac{\sigma_i''(0)}{\kappa_{ij}^{(0)}} - \sigma_j''(0) \right] \nonumber\\
    &= \frac{2}{\left(\alpha^{(1)}_j\right)^2} \left[ \frac{\alpha^{(2)}_i}{\kappa_{ij}^{(0)}} - \alpha^{(2)}_j \right] \equiv \frac{\tilde \epsilon}{\sigma_0}.
\end{align}
We have only kept the first order term in $\delta$ to elucidate how the first correction term arises; one could expand Eq. (\ref{eq:kappa_nonlinear}) to arbitrarily high order if needed.

Armed with (\ref{eq:expansion_kappa}), we can re-work the probability Eq. (\ref{eq:p_anomalous}) and see what the implications are for our statistical model. Including the first order correction to $\kappa_{ij}$, the integrand used in the calculation of the probability of an anomalous failure event now reads:
\begin{align}
    g(\sigma) = W(\sigma; \sigma_0, n) \ S\left(\sigma; \frac{\sigma_0}{\kappa_{ij}^{(0)}(1 + \tilde \epsilon \  \sigma/\sigma_0)}, n\right),
\end{align}
and the probability computes to (discarding all terms higher than $\mathcal{O}(\tilde \epsilon)$, and writing $\kappa_{ij}^{(0)} \equiv \kappa_0$ to declutter the notation):
\begin{align}
    P(s_i>s_j) &\approx \int_0^{\infty} g(\sigma) \mathrm{d}\sigma \nonumber\\
    &= \frac{\kappa_0^n}{1+\kappa_0^n}\left[1 + \tilde \epsilon(\kappa_0) \ w(n, \kappa_0) \right],
\end{align}
where
\begin{align}
    w(n, \kappa_0) = \frac{\left(\kappa_0^n - n\right)}{ \left(1+\kappa_0^n\right)^{1+\frac{1}{n}}} \ \Gamma\left(1+\frac{1}{n}\right).
\end{align}

It is apparent that the correction is positive, which makes sense: if the difference in stress between two elements decreases as $\delta$ increases, the less stressed element is expected to have a higher probability of failure compared to if the stress ratio remained constant. Notice the exponential divergence of the denominator for $n\rightarrow0$---the correction becomes irrelevant as disorder is increased, and at $n=0$ the correction vanishes. We reiterate that this calculation can be repeated up to any order in $\epsilon$, given that the expansion coefficients of the stresses are known, either numerically or analytically.


\section{Evolution of Diffuse Damage}\label{sec:appendix_Evolution_Diffuse}
In the main text, we focus on the probability of diffuse failure at the beginning of the damage process. Herein, we expand on this probability throughout the damage process, and explain why we focus on quantitative estimates at the onset of failure, as well as how estimates are extracted from data.

\subsection{Theory}

We begin by considering the underlying survival distribution of the bulk elements, while they are being progressively and uniformly stressed until a certain small amount of them has failed. Although there are similar discussions in the literature regarding the crack nucleation and emergent statistics of damage evolution starting from a uniform sample~\cite{shekhawat2013damage, bertalan2014fracture}, we are interested in understanding this evolution in the presence of a half-domain crack, which sets a stress scale for the bulk below the point where damage would be expected to become significantly correlated in it~\cite{shekhawat2013damage}. In other words, we assume that no new cracks nucleate and grow far from the pre-existing defect, i.e., that damage in the bulk remains mostly uncorrelated. Our aim is to understand how the amount of diffuse damage evolves, e.g., by finding a relation that connects the proportion of diffuse failure events to the size of the main crack.

Thus, assuming a small amount of failed elements compared to the total number in the \textit{Bulk} set $\mathcal{B}$, i.e., ignoring the correlation between failure events as outlined, we can estimate the amount of bonds that have failed in the bulk as
\begin{equation}
    N_d \approx N_{\mathcal{B}}\left( 1 - e^{(\bar\sigma / \sigma_0)^n} \right),
\end{equation}
where $\bar \sigma$ is the representative bulk stress as in the main text, and $N_{\mathcal{B}}$ is the number of elements in the bulk. Remembering that we have assumed $N_d \ll N_B$, we can make a further approximation:
\begin{equation}
    N_d \approx N_{\mathcal{B}} \left(\bar\sigma/\sigma_0\right)^n.
\end{equation}

With respect to the growth of the main crack cluster (containing $N_f$ fractured sites), to lowest order, we can write it as a function of the crack-tip Weibull stress (which is assumed to relate linearly to the bulk stress, $\sigma_c \propto Y_{\text{\tiny SIF}} \ \bar \sigma$):
\begin{equation}
    N_f \approx v \left(\frac{\sigma_c}{\sigma_0}\right)^\alpha,
\end{equation}
where the exponent $\alpha$ determines how quickly $N_f$ grows (on average and starting from $N_f = 0$) at the initial stages of the damage. We know through our calculations in section \ref{sec:Diffuse_failure}, that the diffuse failure probability $P_d$ at the onset of failure,
\begin{equation}
    P_d = \lim_{\bar \sigma \rightarrow 0} \frac{N_d}{N_d + N_f},
\end{equation}
is non-zero for $n \in \mathbb{R}^+$, i.e., $P_d \in (0,1)$, or equivalently that
\begin{align}
     \lim_{\bar \sigma \rightarrow 0} \left(N_f / N_d\right) \in \mathbb{R^+}.
\end{align}
This is only possible if both $N_f$ and $N_d$ tend to zero equally fast as $\bar \sigma \rightarrow 0$; i.e., if $\alpha = n$. In essence, this guarantees that
\begin{align}
    \lim_{\bar \sigma \rightarrow 0} \left(N_f / N_d\right) &= \frac{v}{N_B} \left(\frac{Y_{\text{\tiny SIF}} \ \bar \sigma}{\sigma_0}\right)^n \left(\frac{\sigma_0}{\bar \sigma}\right)^n \nonumber \\
    &=\frac{v  \ Y^n_{\text{\tiny SIF}}}{N_B} \nonumber \\
    &= 1/\bar \kappa^n \nonumber
\end{align}
which is the form we expect from Eq. (\ref{eq:non_adj_frac}) \& (\ref{eq:k_diffuse}). 

We can then assume that the correction to the expectation value of the ratio $N_f/N_d$ as damage advances, is captured by a generic smooth function $f$:
\begin{equation}
    \frac{N_f}{N_d} = \frac{1}{\bar \kappa^n} \nonumber f\left(\bar \sigma/\sigma_0\right)
\end{equation}
for which $f(0) = 1$. A next to leading order expansion of $f(x) \approx 1+u x^\gamma$, yields a probability $P_d$ that depends on the main crack cluster size as
\begin{equation}
    P_d(N_f) \approx \frac{\bar \kappa^n}{(\bar \kappa^n + 1) + u \ N_f ^{\gamma/n}},
\end{equation}
where $u$ and $\gamma$ are unknown parameters. This result allows us to estimate the probability of diffuse failure at the onset of damage propagation---$P_d(0)$---using simulation data across the failure process, thus increasing the precision of extracted values.

\end{document}